\documentclass{article}

\usepackage{PRIMEarxiv}

\usepackage[utf8]{inputenc} 
\usepackage[T1]{fontenc}    
\usepackage{hyperref}       
\usepackage{url}            
\usepackage{booktabs}       
\usepackage{amsfonts}       
\usepackage{nicefrac}       
\usepackage{microtype}      
\usepackage{lipsum}
\usepackage{fancyhdr}       
\usepackage{graphicx}       
\graphicspath{{media/}}     

\usepackage[figuresright]{rotating}
\usepackage{tablefootnote}
\usepackage{subfig}
\usepackage{graphicx}
\usepackage{multirow}
\usepackage{soul}
\usepackage{enumitem}
\usepackage{adjustbox}
\usepackage{booktabs}

\usepackage{tikz}
\usetikzlibrary{positioning}
\usepackage{pgfplots}
\usetikzlibrary{matrix}

\usepgfplotslibrary{groupplots}
\usepackage{appendix} 
\usepackage{amsmath}

\title{Understanding Diversity in Session-Based Recommendation
\thanks{\textit{\underline{Citation}}: 
\textbf{Qing Yin, Hui Fang, Zhu Sun, Yew-Soon Ong. Understanding Diversity in Session-Based Recommendation. Accepted by TOIS. DOI:10.1145/3600226.}} 
}

\author{
  Qing Yin, Hui Fang \\
  Shanghai University of Finance and Economics \\
  Shanghai, China\\
  \texttt{qyin.es@gmail.com, fang.hui@mail.shufe.edu.cn} \\
   \And
  Zhu Sun \\
  Institute of High Performance Computing; Centre for Frontier AI Research, A*STAR \\
  Singapore\\
  \texttt{sunzhuntu@gmail.com} \\
  \AND
  Yew-Soon Ong \\
  A*STAR Centre for Frontier AI Research; Nanyang Technological University \\
  Singapore \\
  \texttt{asysong@ntu.edu.sg} \\
}

\begin{document}
\maketitle

\begin{abstract}
    Current session-based recommender systems (SBRSs) mainly focus on maximizing recommendation accuracy, while few studies have been devoted to improve diversity beyond accuracy. Meanwhile, it is unclear how the accuracy-oriented SBRSs perform in terms of diversity. Besides, the asserted ``trade-off'' relationship between accuracy and diversity has been increasingly questioned in the literature.
    Towards the aforementioned issues, we conduct a holistic study to particularly examine the recommendation performance of representative SBRSs w.r.t. both accuracy and diversity, striving for better understanding the diversity-related issues for SBRSs and providing guidance on designing diversified SBRSs.
    Particularly, for a fair and thorough comparison, we deliberately select state-of-the-art non-neural, deep neural, and diversified SBRSs, by covering more scenarios with appropriate experimental setups, e.g., representative datasets, evaluation metrics, and hyper-parameter optimization technique. The source code can be obtained via \url{github.com/qyin863/Understanding-Diversity-in-SBRSs}.
    Our empirical results unveil that: 1) non-diversified methods can also obtain satisfying performance on diversity, which can even surpass diversified ones; and 2) the relationship between accuracy and diversity
    is quite complex. Besides the ``trade-off'' relationship,
    they can be positively correlated with each other, that is, having a same-trend (win-win or lose-lose) relationship, which varies across different methods and datasets. Additionally, we further identify three possible influential factors on diversity in SBRSs (i.e., granularity of item
    categorization, session diversity of datasets, and
    length of recommendation lists), and
    offer an intuitive guideline and a potential solution regarding learned item embeddings for more effective session-based recommendation.
\end{abstract}

\maketitle

\section{Introduction}

In recent years,
session-based recommender systems (SBRSs) have {received a lot of attention} for capturing short-term and dynamic user preferences, and thus providing more timely and accurate recommendations, which are sensitive to the evolution of session contexts~\cite{fang2020deep,wang2021survey}.
Existing SBRSs strive to deploy complex models such as deep neural networks to improve the recommendation accuracy by learning a user's short-term preference from the most recent session. For example, GRU4Rec~\cite{hidasi2015session} adopts recurrent neural networks with gated recurrent units (GRU) to capture the sequential behaviors in a session, while NARM~\cite{li2017neural} and STAMP~\cite{LiuZMZ18} further 
adopt attention mechanism to learn a user's main interest (purpose). To capture more complex item relationship, SR-GNN~\cite{WuT0WXT19} and 
GC-SAN~\cite{XuZLSXZFZ19} import graph neural networks (GNNs) based on item graph constructed from the corresponding session to learn more accurate item embeddings. Besides the current session graph, GCE-GNN~\cite{wang2020global} also constructs a global graph from all sessions. 

However, the above popular and representative state-of-the-art SBRSs ignore to consider diversity, which has been recognized, beyond accuracy, as a key factor in satisfying users' diversified demands~\cite{zhang2008avoiding} and promoting enterprises’ sales \cite{liang2021enhancing}.
It is widely known that, those RSs and SBRSs that only seek to improve recommendation accuracy, would lead to overemphasizing dominant interests (e.g. categories) and weakening minor interests for
every user \cite{steck2018calibrated}. More seriously, diversity bias will cause filter bubbles considering the iterative or closed feedback loop in RSs~\cite{nguyen2014exploring,khenissi2020theoretical}.

To this end, diversified RSs aim to provide more diverse recommendation lists, which can be mainly divided into three categories: post-processing heuristic methods~\cite{carbonell1998use,steck2018calibrated}, determinantal point process (DPP) methods~\cite{chen_fast_2018,wu2019pd,gan2020enhancing} and end-to-end learning methods~\cite{zheng2021dgcn,liang2021enhancing}.
However, there are few diversified SBRSs, and to the best of our knowledge, only three representative ones are retrieved, i.e., MCPRN \cite{Wang0WSOC19}, ComiRec \cite{Cen2020ControllableMF} and IDSR \cite{chen2020improving}. MCPRN and ComiRec both assume the existence of multiple purposes instead of only one main purpose in a session, whilst IDSR jointly considers both item relevance and diversity by optimizing on a weighted loss function.
These three diversified SBRSs argue that they have more appropriately involved diversity in contrast to those previous approaches, but \emph{they neglect to moderately compare with other representative SBRSs in terms of accuracy and diversity}. For example, ComiRec is not compared to baselines regarding the diversity metrics, whereas MCPRN merely compares with some deep neural methods w.r.t. both accuracy and diversity metrics, but not with traditional non-neural methods.

Meanwhile, several diversified RSs adopt a ``trade-off'' hyper-parameter~\cite{carbonell1998use, chen_fast_2018, chen2020improving} to combine relevance score and diversification score. Due to such kind of model design, accuracy and diversity are more likely to show accuracy-diversity trade-off in related models. However, \emph{it seems unfair to conclude accuracy-diversity absolute trade-off, as ``common sense'' holds}.
Besides, other studies~\cite{zhou2010solving,zheng2021dgcn} treat accuracy and diversity as conflicting goals, which consistently convey the kind of message that improvements on diversity can only be achieved at the expense of accuracy.
In contrast, there are also some explorations~\cite{wu2019pd} unveiling that, considering diversity adapted to user demands, which are mined from users' historical diversified logs, might facilitate the recommendation performance for both accuracy and diversity.
In this case, \emph{whether there is a trade-off relationship, or others, between accuracy and diversity needs a further thorough exploration}. Moreover, which kinds of factors do lead to diversity difference besides model design, is under-explored.

On the other hand, there are quite a set of surveys on SBRSs~\cite{ludewig2018evaluation, quadrana2018sequence, fang2020deep,wang2021survey}, for the sake of elaborating algorithms and their evaluations, including the measurements of diversity and accuracy. For example, Quadrana et al.~\cite{quadrana2018sequence} propose that evaluations on SBRSs should jointly consider several quality factors (e.g., accuracy and diversity). Ludewig et al.~\cite{ludewig2018evaluation} further compare some SBRSs (e.g., FPMC, GRU4Rec) w.r.t. various measures like accuracy and coverage.
Although previous surveys {empirically} state that a fair and thorough evaluation across different approaches should consider more metrics (e.g., diversity) beyond accuracy, \emph{they ignore to specifically explore the model performance on diversity, and are also lack of a well understanding on the relationship between accuracy and diversity}. Moreover, a fair comparison regarding both accuracy and diversity on the \emph{representative SBRSs}, including ``non-diversified'' (i.e., accuracy-oriented, e.g., NARM and GCE-GNN) and diversified deep neural based SBRSs (e.g., IDSR),
and traditional non-neural methods (e.g., ItemKNN~\cite{sarwar2001item}, FPMC)
are still in the blank.

Towards the aforementioned issues, we conduct a
holistic study to particularly examine the recommendation performance of representative SBRSs
with regard to both accuracy and diversity, aiming to better understand the relationship between accuracy and diversity of different SBRSs across different scenarios, as well as the factors affecting model performance on diversity. 
The main contributions of this work are summarized as follows.
\begin{itemize}[leftmargin=*]
    \item We have thoroughly compared the recommendation performance among state-of-the-art non-diversified and diversified SBRSs on commonly-used datasets across different domains (including e-commerce and music) from accuracy, diversity, and a jointly-considered metric on both accuracy and diversity, which has greatly filled the research gaps in existing surveys.
    Besides, we have also conducted in-depth analysis to disclose the underlying reasons for varied performance on accuracy and diversity regarding different types of deep neural methods.
    \item We have deeply explored the experimental results to check the complex relationship, besides ``trade-off'' one, between accuracy and diversity, from both inter- and intra-model perspectives. 
    \item We have further investigated the influential factors on diversity performance besides the complex model designs, including granularity of item categorization, session diversity of datasets, and {length of recommendation lists}. 
    In addition, we provide a promising suggestion by constraining learned item embeddings for more effective SBRSs. {Furthermore, to help better understand our suggestion, we showcase a potential solution and prove its effectiveness via a demo experiment.}
\end{itemize}

\section{Related work}
Our study is related to {two primary areas: session-based recommendation and diversified studies for traditional and
session-based recommendation scenarios}. The two areas are detailed as below. 
Additionally, we also highlight some concepts that are relevant yet different from our research scope (e.g., Individual Diversity and Fairness).

\subsection{Session-Based Recommendation}\label{subsec:sbr}

The methods on SBRSs can be simply drawn into two groups: traditional non-neural methods and deep neural ones.
Representative traditional methods includes but not limited to
Item-KNN \cite{sarwar2001item}, BPR-MF~\cite{rendle2009bpr}, FPMC~\cite{rendle2010factorizing} and SKNN \cite{jannach2017recurrent}.
Specifically, Item-KNN is an item-to-item method which measures cosine similarity of every two items according to the training set. BPR-MF is a Matrix Factorization (MF) method which optimizes a pairwise ranking loss function via SGD.
FPMC further combines MF with Markov Chain (MC) to better deal with sequential relationship between items. Generally, these methods aim at predicting next actions for users but are not designed especially for session-based scenarios with anonymous users~\cite{ludewig2018evaluation}. 
Besides, they cannot well address the item relationships in relatively longer sequences.
In addition,
compared with Item-KNN, Session-based KNN (SKNN)~\cite{jannach2017recurrent,ludewig2018evaluation} considers session-level similarity instead of only item-level similarity, and thus is capable of capturing best information for more accurate session-based recommendation. 
In particular, for each session, SKNN samples $k$ most similar past sessions in the training data. However, the SKNN does not consider the order of the items
(sequential information) in a session when using the Jaccard index or cosine similarity as the distance measure. Therefore, some SKNN-variants are raised (e.g., V-SKNN~\cite{ludewig2018evaluation} and STAN~\cite{garg2019sequence}) to better consider sequential and temporal
information.

On the contrary, deep neural networks are capable of utilizing a much longer sequence for better prediction \cite{tan2016improved,hidasi2018recurrent}.
For instance, GRU4Rec~\cite{hidasi2015session} is the first to apply recurrent neural network to capture the long-term dependency in a session.
Quite a few extended variants of GRU4Rec have been proposed. For example, Improved GRU4Rec~\cite{tan2016improved} obtains better recommendation performance by designing new data augmentation technique.
Hidasi et al.~\cite{hidasi2018recurrent} design novel ranking loss function and negative sampling method to enhance the effectiveness of GRU4Rec without sacrificing efficiency.
NARM~\cite{li2017neural} further deploys an attention mechanism to model the similarity score between previous items and the last item, and thus captures the main purpose in the session. 
Later, STAMP~\cite{LiuZMZ18} uses simple MLP networks and an attentive net to capture both users’ general interests and current interests. 

However, the above methods always model single-way transitions between consecutive items and neglect the transitions among the contexts (i.e., other items in the session)~\cite{qiu2019rethinking}. To overcome the limitation, GNN-based methods have been designed in recent years. For example, SR-GNN~\cite{WuT0WXT19} imports GNNs to generate more accurate item embedding vectors from the session graph. Similar to SR-GNN, GC-SAN~\cite{XuZLSXZFZ19} replaces the simple attention network with self-attention to capture long-range dependencies by explicitly attending to all the positions. TAGNN~\cite{yu2020tagnn} uses target-aware attention such that the learned session representation vector varies with different target items.
Furthermore, 
GCE-GNN~\cite{wang2020global} learns it over both the current session graph and all-session graph.

It is worth noting that, the aforementioned traditional and deep neural SBRSs are all accuracy-oriented methods, ignoring to consider diversity. This may cause filter bubbles given the iterative or closed feedback loop in RSs~\cite{nguyen2014exploring,khenissi2020theoretical}, thus failing to meet users' diversified demand and decreasing user engagement.

{Next, we particularly summarize the previous surveys on SBRSs, and elaborate the major differences between traditional RSs and SBRSs.}

\subsubsection{{Surveys on SBRSs}}

There are some surveys (including empirical ones) on SBRSs~\cite{ludewig2018evaluation, quadrana2018sequence, fang2020deep,wang2021survey}. 
For example,
Quadrana et al.~\cite{quadrana2018sequence} propose a categorization of recommendation tasks, and discuss approaches for sequence-aware recommender systems where SBRSs is one type of them. Meanwhile, they argue that empirical evaluations should consider multiple quality factors, e.g., accuracy and diversity. 
Ludewig et al.~\cite{ludewig2018evaluation} present an in-depth experimental performance comparison of several SBRSs (FPMC, GRU4Rec, and some simpler methods like Item-KNN)
on evaluation measures (e.g., accuracy and aggregate diversity). 
Fang et al.~\cite{fang2020deep} design a categorization of existing SBRSs in terms of behavioral session types, summarize and empirically demonstrate the key factors affecting the performance of deep neural SBRSs in terms of accuracy-related metrics.
Wang et al.~\cite{wang2021survey} generally define the problems in SBRSs,
and further summarize different data characteristics and challenges of SBRSs.
However, these surveys mainly strive to guide future research by providing an overall picture of existing studies on SBRSs. Although some of them have considered the evaluation issues, or conducted some form of empirical evaluations to compare different models,
\emph{they generally ignore to
specifically explore the model performance on diversity, and thoroughly compare representative SBRSs in terms of both accuracy and diversity}. This, consequently, leads to an insufficient understanding on the relationship between accuracy
and diversity. Our study aims to address these issues with an appropriate empirical design.

\subsubsection{{Remarks on Differences between Traditional RSs and SBRSs}}
{The major differences between traditional recommender systems (RSs) and session-based recommendation systems (SBRSs) lie in three folds.}
{\emph{(1) Different Tasks.} Generally, traditional RSs and SBRSs differ in modeling user preferences. Traditional RSs focus on analyzing all historical interactions of each user to predict her future preferences, while session-based RSs (SBRSs) consider the order and timing of anonymous interactions within a single session to recommend next items for the current session. While traditional RSs are well-established and widely used, SBRSs are recently gaining more and more attention for their ability to handle real-time and personalized recommendations in domains such as e-commerce and music streaming, as SBRSs are more realistic in the real applications.} 
{\emph{(2) Different Techniques.} The difference in user preference modelling consequently leads to different techniques for the two types of RSs. Traditional RSs typically employ, such as matrix factorization techniques, to model the user-item interaction matrix; whereas SBRSs often utilize sequential models such as recurrent neural networks, to capture the sequential patterns hidden in the session, which normally change rapidly.}
{\emph{(3) Different Data Source.} The amount of available data per user also differs between these two types of RSs. In particular, traditional RSs typically possess more data available per user, i.e., all historical interactions of each user. By contrast, SBRSs are better suited for capturing short-term preferences and adapting to user behavior changes in a session data.}

{The inherent differences between the two types of RSs aforementioned directly lead to their differences in diversified modeling. Alternatively stated, the diversified methods designed specifically for RSs cannot be easily transferred into SBRSs due to technical and computational challenges. For instance, it is not practical to exploit the diversified methods (e.g., DPP \cite{chen_fast_2018}) in traditional RSs into SBRSs, as the optimization algorithm in DPP needs to be applied for every session in SBRSs, where the amount of session data is much bigger than the number of users in RSs.}
 
{As SBRSs are gaining increasing attention and more practical in real-world applications, it becomes essential and necessary to provide more diversified recommendations to avoid the filter bubble in the scenarios of SBRSs. It is noteworthy that the research outcomes regarding diversification in traditional RSs and SBRSs are not necessarily contradictory. Our research survey aims to inspire more effective methods for diversified modeling in SBRSs.}

\subsection{Diversified Recommendation}\label{subsec:dr}

Diversity can be viewed at individual or aggregate levels in RSs. Individual diversity depicts the dispersion of recommendation lists, whilst aggregate diversity refers to dispersion from the RS perspective. 
Our paper mainly focuses on individual diversity, that is, 
we explore diversity at individual level if not particularly indicated.

Towards individual diversity in traditional recommendation scenarios,
Carbonell et al. \cite{carbonell1998use} propose the Maximal Marginal Relevance (MMR) to greedily select an item with the local highest combination of similarity score to the query and dissimilarity score to selected documents at earlier ranks. Inspired from dissimilarity score in MMR, some studies \cite{agrawal2009diversifying,santos2010exploiting} define diversification on explicit aspects (categories) or sub-queries. 
Steck \cite{steck2018calibrated} uses the historical interest distribution as calibration to capture minor interests. 
Furthermore, Chen et al. \cite{chen_fast_2018} provide a better relevance-diversity trade-off using DPP in recommendation. The essential characteristic of DPP is that it assigns higher probability to sets of items that are diverse from each other \cite{kulesza2012determinantal}.
Based on the fast greedy inference algorithm~\cite{chen_fast_2018}, some recent studies~\cite{wu2019pd,gan2020enhancing} employ DPP to improve diversity for different recommendation tasks.
However, the above heuristic or DPP-based models are two-stage ones, which consider the diversity in the second stage by re-ranking items ordered by relevance in the first stage. Only several studies~\cite{zheng2021dgcn,liang2021enhancing} are end-to-end ones, that is, jointly optimizing diversity and accuracy by one model. Note that the above studies are for traditional recommendation tasks, rather than anonymous session-based scenarios.

To the best of our knowledge, for session-based recommendation, there are only three end-to-end diversified works, i.e.,  MCPRN~\cite{Wang0WSOC19}, ComiRec~\cite{Cen2020ControllableMF}, and IDSR~\cite{chen2020improving}.
In particular, MCPRN uses mixture-channel purpose routing networks to guide the multi-purpose learning, while ComiRec explores two methods, namely dynamic routing method and self-attentive method, as multi-interest extraction module. MCPRN and ComiRec both use multiple session representations to capture diversified preferences, which can implicitly satisfy diversified user demand.
On the contrary, 
IDSR explicitly constructs set diversity and achieves end-to-end recommendation guided by the intent-aware diversity promoting (IDP) loss. The final user preference towards an item is a combination of the relevance score and diversification score, 
weighted by a ``trade-off hyper-parameter'' (as defined in IDSR) controlling the balance between accuracy and diversity.
However, as we have discussed, whether there is other kind of relationship, in addition to trade-off, between accuracy and diversity, and how diversified methods perform compared to other non-diversified SBRSs on diversity, are both under-explored.

\subsection{Discussions on Relevant yet Different Concepts}

\subsubsection{Session-Based Recommender Systems (SBRSs) vs. Sequential Recommender Systems (SRSs)}
SBRSs and SRSs are built on session data and sequence data, respectively, while they are often
gotten mixed up by some readers since both of them consider the sequential information of interactions~\cite{wang2021survey}.
In academia, SRSs are typically operationalized as the task of predicting the next user action~\cite{ludewig2018evaluation} based on user sequence data. That is, 
a single sequence data contains all historical, time-ordered logs for a given user, e.g., his/her item viewing and purchase activities on an e-commerce shop, or  listening history on a music streaming site.
On the contrary, as we have defined, SBRSs commonly consider anonymous sessions where user information (including identities) is unknown.
Besides, different from the relatively longer user sequences in SRSs, a session usually contains fewer interactions and is bounded in a shorter time window~\cite{wang2021survey}, e.g., one-day~\cite{ludewig2018evaluation} or 30-minutes ~\cite{luo2020collaborative}.
Our paper focuses on session-based recommendation which recommends the Top-$N$ list for next-item prediction. Therefore, some popular SRSs (e.g., SASRec~\cite{kang2018self} and BERT4Rec~\cite{sun2019bert4rec}) are out of our research scope, and thus are not included in Baselines.

\subsubsection{Individual Diversity vs. Aggregate Diversity}
Diversity can be viewed at individual or aggregate levels in RSs. Specifically, individual diversity depicts the dispersion of recommendation lists, whilst aggregate diversity refers to dispersion from the RS perspective~\cite{wu2019recent}. That is to say, the individual diversity is for recommended items to each individual user regardless of other users, but the aggregate diversity is for all recommended items across all users, which mainly considers overall product variety and sales concentration (e.g., Coverage~\cite{kim2019sequential}). For example, some studies~\cite{kim2019sequential,liu2020long} utilize long-tail (less popular or frequent) items to further improve aggregate diversity. Besides, some works~\cite{gupta2019niser, gupta2021causer} dealing with popularity bias also contribute to higher aggregate diversity.
Importantly, 
aggregate and individual diversity are not necessarily correlated. For example, a system can recommend the same set of highly diverse items to everyone and thus obtains higher individual diversity, which does not lead to high aggregate diversity.
Our paper focuses on individual diversity, that is, we explore diversity at individual level if not particularly indicated.

\subsubsection{Diversity vs. Fairness.}
{Fairness in recommendations is built on notions of inclusion, non-discrimination, and justice~\cite{kleinberg2016inherent, schelenz2021diversity}. The fairness-related studies are two-fold: item fairness (i.e., same probability of being displayed between items with the same value on attributes)~\cite{kleinberg2016inherent, wang2021user, rahmani2022experiments} and user fairness (i.e., treat different user groups similarly)~\cite{leonhardt2018user, sacharidis2020fairness, wang2021user, li2021user, rahmani2022experiments}. Fairness and diversity are intertwined in several ways. For instance, in order to increase aggregate diversity (e.g., item coverage), RSs frequently encourage long-tail item exposure. The same way also can improve item fairness. Although individual and aggregate diversity can be used to increase fairness for users and items respectively, they do not address other aspects of fairness, such as statistical parity~\cite{drosou2017diversity} and differential treatment of two users or two items~\cite{leonhardt2018user}. For example, there are $10\%$ women and $90\%$ men among job applicants. While RSs with diversity objectives hope for a uniform gender recommendation distribution (i.e., $50\%$ for women and $50\%$ for men), statistical parity requires that the distribution of results across genders be the same as the whole population. It is not always the case that fairness and diversity objectives are in agreement (e.g., diversity improving algorithms can lead to discrimination among users~\cite{leonhardt2018user}). Our paper focuses on the accuracy and individual diversity performance and explores their relationship in SBRSs.}

\section{Experimental Settings}
Instead of only evaluating recommendation accuracy, we further explore diversity in the session-based scenario.
Towards a fair study to draw convincing results, we aim to cover more scenarios with appropriate experimental setups. Specifically, we select representative datasets across different domains, including e-commerce and music (Section~\ref{subsec:datasets}); 
we moderately choose three types of session-based methods for comparison, namely traditional non-neural methods, state-of-the-art deep neural methods, and {three} diversified ones (Section~\ref{subsec:baselines}); and we take a comprehensive set of accuracy- and diversity-related indicators (Section \ref{subsec:metrics}).
Accordingly, we conduct extensive experiments to answer three key research questions (RQs):
%
\begin{itemize}[leftmargin=*]
    \item \textbf{RQ1}: How do representative SBRSs of different types perform in terms of accuracy and diversity metrics? 
    Furthermore, what are the possible reasons leading to their varied performance?
    \item \textbf{RQ2}: Whether there is a ``trade-off'' relationship between accuracy and diversity? Is there any other one between them? 
    \item \textbf{RQ3}: Which kinds of factors will influence diversity performance of SBRSs, besides various model designs?
    
\end{itemize}
\subsection{Datasets and Preprocessing}\label{subsec:datasets}
We delicately select four representative public datasets for the experimental purpose.
They are three e-commerce datasets (i.e., Diginetica\footnote{\url{https://competitions.codalab.org/competitions/11161\#learn\_the\_details-overview}.}, Retailrocket\footnote{\url{https://www.kaggle.com/retailrocket/ecommerce-dataset}.}, Tmall\footnote{\url{https://tianchi.aliyun.com/dataset/dataDetail?dataId=42}.}) with item category information and one music dataset (i.e., Nowplaying\footnote{\url{https://zenodo.org/record/2594483\#.YdfMgxNBy8U}.}) with artist information.

\begin{itemize}[leftmargin=*]
    \item \textbf{Diginetica} comes from CIKM Cup 2016 and includes user e-commerce search engine sessions with its own `SessionId'. Note that we only use the `view’ data.
    \item \textbf{Retailrocket} collects users' interaction behavior in the e-commerce website over $4.5$ months. We also only explore interactions of `view' type, and partition user history into sessions in every $30$-minute interval following~\cite{luo2020collaborative}.
    \item \textbf{Tmall} comes from the IJCAI-15 competition and contains anonymous shopping logs on Tmall. We adopt interactions of `buy' and `view' action-types, and partition user history into sessions by day following~\cite{ludewig2018evaluation}.
    Since the original datasets are quite large, we select $1/16$ sessions as sampling inspired by fractions of Yoochoose~\cite{li2017neural}.
    \item \textbf{Nowplaying} tracks users' current listening from music-related tweets. Ludewig et al. \cite{ludewig2018evaluation} publicize the processed version\footnote{\url{www.dropbox.com/sh/dbzmtq4zhzbj5o9/AACldzQWbw-igKjcPTBI6ZPAa?dl=0}.} with `SessionId'. We use `ArtistId' as the category information to distinguish different music for simplicity by following the previous studies (e.g., \cite{zhu2020sequential} uses region information to represent category on the POI dataset Gowalla).
\end{itemize}

\begin{table}[t]
    \centering
    \caption{Statistics of Datasets (Note: \# train and \#test represent the number of sessions before sequence splitting preprocess).}
    \begin{tabular}{lrrrr}
    \toprule
    Dataset &  Diginetica & Retailrocket & Tmall & Nowplaying\\
    \midrule
    \# interactions & 993,483 & 1,082,246 & 1,505,683  & 1,227,583\\
    \# train & 186,670 & 294,629 & 188,756   & 144,356\\
    \# test &  18,101 & 12,206 & 51,894  & 1,680\\
    \# items &  43,097 & 48,893 & 96,182  & 60,622\\
    \# categories & 995 & 944 &  822  & 11,558\\
    avg. len. & 4.8504 & 3.5253 & 6.0775  & 8.4056\\
    \bottomrule
    \end{tabular}
    \label{tab:datasetStatistics}
\end{table}

For data preprocessing, following~\cite{li2017neural,LiuZMZ18,WuT0WXT19}, to filter noisy data, we drop sessions of length $1$ and items occuring less than $5$ times. We set the most recent data (a week) as the test set whilst the other sessions as the training set. Besides, we further filter out items appearing in the test set but not in the training set. The statistics of these four datasets after preprocessing are shown in Table~\ref{tab:datasetStatistics}.
It should be noted that sequence splitting preprocess \cite{li2017neural} is necessary if a recommendation model is not trained in session-parallel manner \cite{hidasi2015session}. Sequence splitting preprocess refers to that, for a session sequence $S=[i_1, i_2, \dots, i_n]$, we can generate $n-1$ sub-sequences $([i_1], i_2)$, $([i_1, i_2], i_3)$, $\dots$, $([i_1,\dots, i_{n-1}], i_n)$ for training.

\subsection{Baseline Models}\label{subsec:baselines}
To explore the recommendation performance on accuracy and diversity, we select three categories of popular and representative baseline models for session-based recommendation: \emph{traditional non-neural methods}, \emph{deep neural methods}, and \emph{deep diversified methods}.

\subsubsection{Traditional Non-neural Methods}
\begin{itemize}[leftmargin=*]
    \item \textbf{POP} always recommends top ranking items based on popularity in the training set.
    \item \textbf{S-POP} recommends top frequent items of the current session, which differs from \text{POP} using global popularity values.
    Ties are broken up using global popularity values.
    \item \textbf{Item-KNN}~\cite{sarwar2001item}
    is an item-to-item model which measures cosine similarity of every two items regarding sessions in the training data. For a session, it
    recommends the most similar items to the last item of the session.
    \item \textbf{BPR-MF}~\cite{rendle2009bpr} optimizes a pairwise ranking loss on Matrix Factorization (MF) method.
    It further averages items' feature vectors in the session as its feature vector.
    
    \item \textbf{FPMC}~\cite{rendle2010factorizing} is a sequential method based on MF and first-order MC. To adapt to anonymous session-based recommendation, it drops user latent representations.
\end{itemize}

\subsubsection{Deep Neural Methods}

\begin{itemize}[leftmargin=*]
    \item \textbf{GRU4Rec}~\cite{hidasi2015session} is an RNN-based model which utilizes session-parallel mini-batch training process and also adopts pairwise ranking loss function.

    \item \textbf{NARM}~\cite{li2017neural} is an RNN-based model with an attention mechanism to capture the main purpose from the hidden states and combine it with the last hidden vector as the final representation to generate recommendations.
    
    \item \textbf{STAMP}~\cite{LiuZMZ18} employs attention layers directly on item representation instead of the output of RNN encoder then captures the user’s long-term preference from session context, and the short-term interest according to a session's last item.
    \item \textbf{SR-GNN}~\cite{WuT0WXT19} employs a gated GNN layer to obtain item embeddings and then applies an attention mechanism to compute the session representations.
    
    \item \textbf{GC-SAN}~\cite{XuZLSXZFZ19} is quite similar to SR-GNN, except it uses Self-Attention Network (SAN) to learn session representations.
    \item \textbf{GCE-GNN}~\cite{wang2020global} constructs both current session (local) graph and global graph to get session- and global-level item embeddings. Then, position-aware attention is adopted to fuse reversed position information to obtain the final session representation.
\end{itemize}

\subsubsection{Deep Diversified Methods}

\begin{itemize}[leftmargin=*]
    \item \textbf{MCPRN}~\cite{Wang0WSOC19} models users’ multiple purposes (instead of only main purpose as NARM) in a session. It further uses target-aware attention to combine those learned multiple purposes to get the final representation. As claimed in the original paper, MCPRN can boost both accuracy and diversity.
    \item \textbf{NARM+MMR}~\cite{chen2020improving} {is a two-stage approach which in the second stage uses MMR~\cite{carbonell1998use} and a greedy algorithm to re-rank items provided by NARM in terms of relevance scores in the first stage.} 
    \item \textbf{IDSR}~\cite{chen2020improving} is the first end-to-end deep neural network based method that jointly considers diversity and accuracy for SBRSs. It presents a novel loss function to guide model training in terms of both accuracy and diversity, where hyper-parameter $\lambda$ is adopted to balance the relevance score and diversification score.
    
\end{itemize}

\subsection{Evaluation Metrics}\label{subsec:metrics}
For an exhaustive evaluation, we adopt the following metrics related to accuracy, diversity, or both. A higher value of each metric indicates better performance. 
Specifically, to evaluate accuracy~\cite{li2017neural,WuT0WXT19,wang2020global}, we adopt HR (Hit Rate), MRR (Mean Reciprocal Rank), and NDCG (Normalized Discounted Cumulative Gain). In particular, \textbf{HR} measures whether a ground-truth item is contained in the Top-$N$ Recommended List (abbreviated as RL, and $N$ is the length of the RL);
\textbf{MRR} measures whether a correctly predicted item ranks ahead in the RL;
and \textbf{NDCG} rewards each hit based on its position in the RL. 

Towards diversity, 
we choose ILD (Intra-List Distance)~\cite{hu2017diversifying,Cen2020ControllableMF,chen2020improving}, Entropy~\cite{Wang0WSOC19,zheng2021dgcn}, and Diversity Score~\cite{liang2021enhancing}. To be specific, \textbf{ILD} measures the average distance between every pair of items in RL where $d_{ij}$ denotes the euclidean distance between the respective embeddings (e.g., one-hot encoding) of categories that items $i$ and $j$ belong to,
\begin{equation}
        \text{ILD} = \frac{\sum_{(i,j)\in RL}d_{ij}}{|RL|\times(|RL|-1)};\label{eq:ild}
\end{equation}
\textbf{Entropy} measures the entropy of item category distribution in the RL. The more dispersed category distribution is, the more diverse the RL is; and
\textbf{Diversity Score} (shorted as \textbf{DS}) is calculated by the number of interacted/recommended categories divided by number of interacted/recommended items.

Furthermore, we adopt \textbf{F-score}~\cite{hu2017diversifying} as an aggregative indicator which jointly considers both accuracy and diversity. Here, F-score is computed as the Harmonic mean of accuracy metric (i.e., HR) and diversity metric (i.e., ILD),
\begin{equation}
\text{F-score}=\frac{2 \text{HR}\times \text{ILD}}{\text{ HR}+\text{ILD}}\label{eq:F-score}.
\end{equation}
A higher F-score implies that the corresponding model has a more comprehensive strength with regard to both accuracy and diversity.

\subsection{Hyper-parameters Setup}
For deep models, we use the Adam optimizer. We tune hyper-parameters of all baseline models on a validation set which is the most recent data (last week) of every training set.

Noted that different baselines have different hyper-parameters, where the most common ones include item embedding dimension, dimension of latent vector, learning rate, the size of mini-batch, and the number of epochs. 
Considering a fair comparison, we use the Bayesian TPE~\cite{bergstra2011algorithms} of Hyperopt\footnote{\url{https://github.com/hyperopt/hyperopt}.} framework to tune all hyper-parameters of baselines on all datasets, which has proven to be a more intelligent and effective technique compared to grid and random search, especially for deep methods (having more hyper-parameters)~\cite{sun2020we}.
\begin{table*}[]

\centering
\caption{The Optimal Hyper-parameter Settings by Bayesian TPE of Hyperopt.}
\begin{adjustbox}{max width=\textwidth}
\begin{tabular}{l|l|c|c|c|c|l|l}
\toprule
Model                   & Hyper-parameter       &  Digi* & Retail* & Tmall & Now* & Searching Space             & Description \\\midrule
Item-KNN                & -alpha                &0.9270            & 0.7100           & 0.8514             &0.9074       & $\mathcal{U}(0.1,1)$           &Balance for normalizing items' supports             \\\hline
\multirow{4}{*}{BPR-MF} & -item\_*\_dim &300            & 100           &200              &150       & $[min=100,max=300,step=50]$       & the dimension of item embedding             \\
                        & -lr                   &0.01            & 0.01           &0.001              &0.001       & $[0.001,0.005,0.01,0.05]$ & learning rate             \\
                        & -batch\_size          &64           &64            &512              &512       & $[64,128,256,512]$        & the size for mini-batch            \\
                        & -epochs               &20            & 20           &40              & 15      & $[min=10,max=40,step=5]$         &the number of epochs             \\\hline 
\multirow{4}{*}{FPMC} & -item\_*\_dim &250            &100            &200              & 250      & $[min=100,max=300,step=50]$       &             \\
                        & -lr                   & 0.005           &0.001            &0.001              &0.005       & $[0.001,0.005,0.01,0.05]$ &             \\
                        & -batch\_size          & 256           &256           &512              &512       & $[64,128,256,512]$        &             \\
                        & -epochs               & 30           &10            & 40             &40      & $[min=10,max=40,step=5]$         &             \\\hline 
\multirow{8}{*}{GRU4Rec} & -item\_*\_dim &150            &300            &300              &100       & $[min=100,max=300,step=50]$       &             \\
                        & -lr                   &0.05            &0.01            &0.05              &0.01      & $[0.001,0.005,0.01,0.05]$ &             \\
                        & -batch\_size          &256            &256           &64              &512       & $[64,128,256,512]$        &             \\
                        & -epochs               &25            &30            &30              &35      & $[min=10,max=40,step=5]$         &             \\
                        & -hidden\_size               &50            &200            &150              &200       & $[min=50,max=200,step=50]$          & the dimension of latent vector             \\
                        & -n\_layers               &1            &1            &1              &1       & $[1,2,3]$         & the number of layers in RNN            \\
                        & -dropout\_input               &0.3123            &0.1073            &0.3697              &0.1407       & $\mathcal{U}(0.1,1)$          & dropout rate             \\
                        & -dropout\_hidden               &0.2946            &0.1311            &0.6618             &0.4170       & $\mathcal{U}(0.1,1)$          &dropout rate             \\\hline  
\multirow{6}{*}{NARM} & -item\_*\_dim &200            &100            &250              &150       & $[min=100,max=300,step=50]$       &             \\
                        & -lr                   &0.001            &0.001            &0.005              &0.001       & $[0.001,0.005,0.01,0.05]$ &             \\
                        & -batch\_size          &512           &512            &256              &512       & $[64,128,256,512]$        &             \\
                        & -epochs               &35            &40            &25              &10       & $[min=10,max=40,step=5]$         &             \\
                        & -hidden\_size               &50            &150            &150              & 150      & $[min=50,max=200,step=50]$          &             \\
                        & -n\_layers               &1            &1            &1              &2       & $[1,2,3]$         &             \\\hline 
\multirow{4}{*}{STAMP} & -item\_*\_dim &100            &100            &150              &200       & $[min=100,max=300,step=50]$       &             \\
                        & -lr                   &0.001            &0.001            &0.01              &0.001       & $[0.001,0.005,0.01,0.05]$ &             \\
                        & -batch\_size          &128            &512            &256             &128       & $[64,128,256,512]$        &             \\
                        & -epochs               &35            &20            &35              &15       & $[min=10,max=40,step=5]$         &             \\\hline                        
\multirow{5}{*}{SR-GNN} & -item\_*\_dim &300            &150            &150              &200       & $[min=100,max=300,step=50]$       &             \\
                        & -lr                   &0.005            &0.005            &0.005              &0.005       & $[0.001,0.005,0.01,0.05]$ &             \\
                        & -batch\_size          &256            &256            &128              &512       & $[64,128,256,512]$        &             \\
                        & -epochs               &25            &20            &15             &10       & $[min=10,max=40,step=5]$         &             \\
                        & -step              &1            &1            &3              &3       & $[1,2,3]$          &gnn propogation steps             \\\hline
\multirow{6}{*}{GC-SAN} & -item\_*\_dim & 250           &150            &150              &300       & $[min=100,max=300,step=50]$       &             \\
                        & -lr                   &0.001            &0.001            &0.005              &0.001       & $[0.001,0.005,0.01,0.05]$ &             \\
                        & -batch\_size          &512            &512            &256              &256       & $[64,128,256,512]$        &             \\
                        & -epochs               &25            &40            &10              &30       & $[min=10,max=40,step=5]$         &             \\
                        & -weight               &0.4            &0.4            &0.4              &0.4       & $[0.4,0.6,0.8]$          & weight factor (in combined embedding)            \\
                        & -blocks               &3            &4           &1              &3       & $[1,2,3,4]$        &the number of
stacked self-attention blocks              \\\hline   
\multirow{7}{*}{GCE-GNN} & -item\_*\_dim &250            &100            &100              &100       & $[100]$        &             \\
                        & -lr                   &0.001            &0.001            &0.005              &0.001       & $[0.001,0.005]$ &             \\
                        & -batch\_size          &128           &100            &100              &100       & $[100]$        &             \\
                        & -epochs               &10            &30            &20              &30       & $[min=10,max=30,step=5]$          &             \\
                        & -n\_iter               &1            &1            &2              &1       & $[1,2]$          & the number of hop             \\
                        & -dropout\_gcn               &0.4            &0.4            &0.2              &0.0       & $[0,0.2,0.4,0.6,0.8]$          &dropout rate             \\                
                        & -dropout\_local               &0.5            &0.0            &0.0              &0.0       & $[0,0.5]$         &dropout rate             \\\hline 
\multirow{6}{*}{MCPRN} & -item\_*\_dim &150            &150            &100              &200       & $[min=100,max=200,step=50]$        & dimension of item embedding/latent vector              \\
                        & -lr                   &0.005            &0.005            &0.005              &0.005       & $[0.005,0.01,0.05]$ &             \\
                        & -batch\_size          &256            &50            &50              &50       & $[50]$        &             \\
                        & -epochs               &15            &30            &25              &15       & $[min=10,max=40,step=5]$         &             \\        
                        & -tau               &1            &0.01            &0.01              &0.1       & $[0.01,0.1,1,10]$          & temperature parameter in softmax             \\ 
                        & -purposes               &1            &4            &1              &3       & $[1,2,3,4]$          &The number of channels             \\ \hline
\multirow{4}{*}{Remark}  & \multicolumn{7}{|l}{1. Digi* represents Diginetica, Retail* for Retailrocket, Now* for Nowplaying, item\_*\_dim for item\_embedding\_dim.}\\
                         & \multicolumn{7}{|l}{2. Omit the hyper-parameter description if exists before.}\\
                         & \multicolumn{7}{|l}{3. Due to memory limit, set item\_*\_dim, batch\_size as 100 (original setting) in GCE-GNN, and batch\_size as 50 in MCPRN except Digi*.}\\
                         & \multicolumn{7}{|l}{4. IDSR uses own official TensorFlow code with early-stopping. Tune $\lambda_e \in [0.1,1]$ and set it as 1 for four datasets.}\\\bottomrule
\end{tabular}
\end{adjustbox}
\label{tab:optimalhypers}
\end{table*}
The detailed optimal hyper-parameter settings by Hyperopt of the baselines are shown in Table \ref{tab:optimalhypers}.
The exceptions are made on that we set both item embedding dimension and mini-batch size as $100$ (consistent with the original paper setting) for GCE-GNN due to memory space limits. Similarly, we set the mini-batch size as $50$ for MCPRN.

\begin{figure}[t]
    \centering
    \includegraphics[scale=0.32]{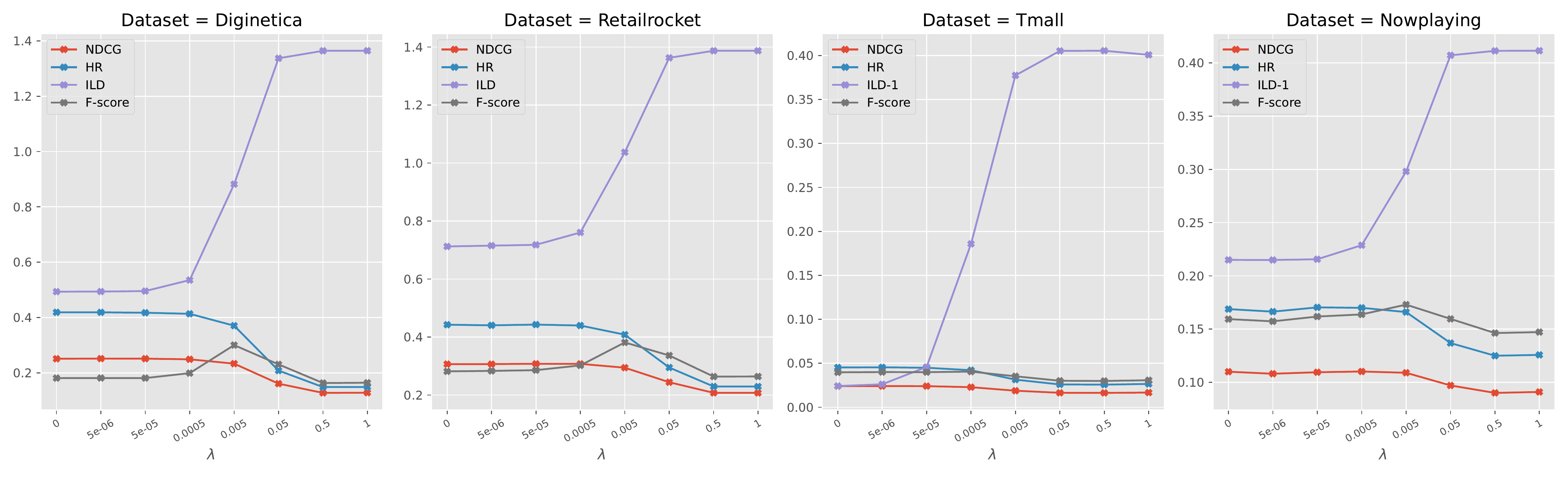}
    \caption{{The Impact of MMR for NARM+{MMR} with $N=10$.}}
    \label{fig:mmr_fined}
\end{figure}

{
Besides, the IDSR approach is an end-to-end method that aims to balance and improve accuracy and diversity simultaneously. This is achieved through the use of a hyper-parameter $\lambda$ in the range of $[0,1]$ with the formula $\lambda*$relevance score $+(1-\lambda)*$ diversification score. To evaluate the performance of IDSR, we set $\lambda$ to {0.2, 0.5, 0.8} for each dataset. These three variants of IDSR allow us to assess the impact of $\lambda$ on the overall effectiveness of IDSR.
Additionally, the NARM+MMR approach involves a two-stage process in which items are reranked based on a fixed relevance score from NARM using the formula $\text{relevance score}+\lambda * \text{diversification score}$ , where $\lambda$ is a multiplier chosen from a range of values, including $\{0, 5e-6, 5e-5, 5e-4, 0.005, 0.05, 0.5, 1\}$.
Using the principle of maximum F-score searching, as shown in Figure~\ref{fig:mmr_fined}, we determined the optimal value of $\lambda$ for each dataset (i.e., $\lambda=5e-4$ for the Tmall dataset and $\lambda=0.005$ for the other three datasets).
}

We have integrated all the codes with PyTorch framework, except for IDSR that we adopt its original code in TensorFlow version\footnote{\url{https://bitbucket.org/WanyuChen/idsr/}.} with an early-stopping mechanism. 
The source code and datasets are available online\footnote{\url{https://github.com/qyin863/Understanding-Diversity-in-SBRSs}.}.

\section{Experimental Results}
\label{sec:results}
In this section, we present our experimental results to answer the raised three research questions (RQs). {Besides, to enhance clarity, we include a condensed overview of significant findings, presented as Table~\ref{tab:summary}.}
\begin{table}[]
\newcommand{\tabincell}[2]{
\begin{tabular}{@{}#1@{}}#2\end{tabular}
}
\caption{Summary of Important Findings in Section \ref{sec:results}.}\label{tab:summary}
\begin{adjustbox}{width=1\linewidth}
\begin{tabular}{l|l|l}\toprule
Topic& Important Findings& Support \\\midrule
\tabincell{c}{Accuracy\\Performance} & \begin{tabular}[c]{@{}l@{}}Deep neural methods generally outperform traditional non-neural methods.\\ Deep diversified methods perform worse than non-diversified deep methods, \\but better than traditional ones.\end{tabular}                              & Tables \ref{tab:bordacount}-\ref{tab:nowplaying}          \\\hline
\tabincell{c}{Diversity\\Performance}            & \begin{tabular}[c]{@{}l@{}}POP, S-POP, and GRU4Rec, though not being specifically designed for diversity purpose,\\ perform comparably to IDSR.\end{tabular}                                                                                                                                                                                             & Tables \ref{tab:bordacount}-\ref{tab:nowplaying}                                                                                                                                                                          \\\hline
\tabincell{c}{Comprehensive\\ Performance}        & \begin{tabular}[c]{@{}l@{}}Deep neural method obtain better comprehensive performance than non-neural traditional methods.\\ Accuracy-oriented approaches (e.g. GC-SAN, STAMP), although not particularly designed \\for diversity purpose, can achieve a satisfying balance between accuracy and diversity.\end{tabular}                                                                      & Tables \ref{tab:bordacount}-\ref{tab:nowplaying}                                                                                                                                                                          \\\hline
\tabincell{c}{In-depth Analysis\\(Case Study)}           & \begin{tabular}[c]{@{}l@{}}The learned item embeddings of different categories by GRU4Rec are mixed and inseparable, \\accounting for the highest diversity and lowest accuracy.\\ Attention mechanism can well address noisy information and distinguish item information.\end{tabular}                                                 & Figures \ref{fig:umap_digi_20},\ref{fig:umap_digi_noatt_20}                                                                                                                                                                         \\\hline
\tabincell{c}{Accuracy-Diversity\\Relationship}  & \begin{tabular}[c]{@{}l@{}} The relationship between accuracy and diversity is quite complex and mixed.\\Beside the “trade-off” relationship, they can be positively correlated with each other.\\ The model design and characteristics of datasets can lead to the mixed relationship.\end{tabular}                                                                                                                          & \begin{tabular}[c]{@{}l@{}}Inter- (Tables \ref{tab:bordacount}-\ref{tab:nowplaying}, Figures \ref{fig:hr_ild_scatter}, \ref{fig:corr_by_da})\\intra-model (Figures \ref{fig:corr_by_md}, \ref{fig:corr_by_md_da})\\ Case explanation\end{tabular} \\\hline
\tabincell{c}{Influential Factors\\of Diversity} & \begin{tabular}[c]{@{}l@{}}The granularity of item categorization and the length of lists cause potential biases for diversity indicators.\\ The diversity performance is positively correlated with the session diversity of both input datasets and test samples.\\ We present a potential direction for diversified SBRSs inspired by embedding distribution and \\propose a potential solution through a demo experiment.\end{tabular} & \begin{tabular}[c]{@{}l@{}}Figure \ref{fig:lenbias}\\ Figures \ref{fig:dsinfluential}, \ref{fig:dsimpact_test}\\Figures \ref{fig:cpd_result_10}, \ref{fig:stamp_umap_ecom}\\ \end{tabular}\\\bottomrule
\end{tabular}
\end{adjustbox}
\end{table}

\subsection{Overall Comparisons (RQ1)}
Experimental results of the selected baselines on the four real-world datasets are respectively presented in Tables~\ref{tab:digi}-\ref{tab:nowplaying}, where the best result under each metric is highlighted in boldface and the runner-up is underlined\footnote{We can get similar results concerning baselines of different scenarios with $N=5$, which are not mentioned in the main paper.}. The best results are marked with `$*$' which demonstrates it outperforms the runner-up at
$95\%$
confidence level in t-test for means
of paired two sample.
Note that the results are measured as an average of $5$ times with the best hyper-parameter settings.
{Additionally, we plot the results from Tables~\ref{tab:digi}-\ref{tab:nowplaying} to more clearly display relationship between accuracy and diversity.}
We further adopt a Borda count~\cite{emerson2013original} ranked voting scheme to aggregate our experimental results on the four datasets, for better overall comparisons. Specifically, for these baselines ({16} in all), on each dataset in terms of each metric regarding a specific Top-$N$ recommendation list ($N=\{5, 10, 20\}$), the first-ranked one receives {15} points, and the last-ranked one gains 0 points. We consider every such scenario as a vote. For each baseline, we aggregate all ranking points regarding accuracy metrics (NDCG, MRR, and HR) as Accuracy points, while those for diversity metrics (ILD, Entropy and DS) as Diversity points. We then rank these baselines regarding Accuracy, Diversity and F-score points, respectively, considering that more points refer to better performance. The results are shown in Table \ref{tab:bordacount}.

\begin{table}[t]
    \centering
    \caption{Borda Count and Corresponding Rank of Baselines On Accuracy, Diversity and F-score.}
    \begin{tabular}{l|cc|cc|cc}
    \toprule
{} &  \textbf{Accuracy} &Rank&  \textbf{Diversity} &Rank &  \textbf{F-score} &Rank \\
\midrule
POP       &   10 &        16 &  331 &         3 &        5 &      16 \\
S-POP     &  133 &        11 &  315 &         5 &       78 &       6 \\
Item-KNN  &  195 &         8 &   69 &        14 &       50 &      13 \\
BPR-MF    &   96 &        14 &  176 &         9 &       37 &      14 \\
FPMC      &   42 &        15 &  192 &         8 &       16 &      15 \\\hline
GRU4Rec   &  124 &        13 &  331 &         4 &       61 &      12 \\
NARM      &  359 &         2 &   43 &        15 &       76 &       8 \\
STAMP     &  313 &         4 &  129 &        11 &       85 &       5 \\
SR-GNN    &  290 &         5 &   91 &        13 &       73 &       9 \\
GC-SAN    &  330 &         3 &  113 &        12 &       89 &       3 \\
GCE-GNN   &  397 &         1 &    1 &        16 &       65 &      10 \\\hline
MCPRN     &  199 &         7 &  176 &        10 &       62 &      11 \\
NARM+MMR  &  258 &         6 &  288 &         6 &      121 &       1 \\
IDSR($\lambda=0.2$) &  125 &        12 &  387 &         1 &       88 &       4 \\
IDSR($\lambda=0.5$) &  182 &        10 &  340 &         2 &       96 &       2 \\
IDSR($\lambda=0.8$) &  187 &         9 &  258 &         7 &       78 &       7 \\
    \bottomrule
    \end{tabular}
    \label{tab:bordacount}
\end{table}

\setlength\rotFPtop{0pt plus 1fil} 
\begin{sidewaystable}[htbp]
\footnotesize
\caption{Model Performance on Diginetica. $*$ denotes the best model significantly outperforms the runner-up using a paired t-test (p-value $\leq 0.05$).}\label{tab:digi}
\centering
    \begin{tabular}{c|cc|cc|cc|cc|cc|cc|cc}
        \toprule
         Model & \multicolumn{2}{c}{NDCG} & \multicolumn{2}{c}{MRR} & \multicolumn{2}{c|}{HR} & \multicolumn{2}{c}{ILD} & \multicolumn{2}{c}{Entropy} & \multicolumn{2}{c|}{DS} & \multicolumn{2}{c}{F-score}\\\hline
         & @10 & @20& @10 & @20& @10 & @20& @10 & @20& @10 & @20& @10 & @20& @10 & @20\\\midrule
         POP & 0.0025 & 0.0027 & 0.0017 & 0.0018 & 0.0052 & 0.0061 & \underline{1.1314} & \underline{1.2256} & 2.0464 & \underline{2.6826} & 0.5000 & \underline{0.5333} & 0.0055 & 0.0067\\ 
         S-POP &0.1625 & 0.1630 & 0.1475 & 0.1476 & 0.2083 & 0.2100 & 1.1147 & 1.1914 & \underline{2.1088} & 2.5454 & 0.5326 & 0.4891 & 0.2152 & 0.2274\\ 
         Item-KNN & 0.1313 & 0.1438 & 0.0999 & 0.1036 & 0.2343 & 0.2814 & 0.1653 & 0.2247 & 0.2852 & 0.4353 & 0.1562 & 0.1376 & 0.0375 & 0.0635\\ 
         BPR-MF & 0.0799 & 0.0954 & 0.0618 & 0.0661 & 0.1397 & 0.2012 & 0.5334 & 0.5799 & 0.9490 & 1.2148 & 0.2871 & 0.2159 & 0.0676 & 0.1061\\ 
         FPMC & 0.0787 & 0.0954 & 0.0592 & 0.0638 & 0.1428 & 0.2094 & 0.5566 & 0.5968 & 1.0143 & 1.2854 & 0.3035 & 0.2282 & 0.0554 & 0.0904\\\hline
         GRU4Rec & 0.1220 & 0.1396 & 0.0938 & 0.0987 & 0.2141 & 0.2837 & 0.8351 & 0.9126 & 1.7810 & 2.4512 & \underline{0.5406} & 0.5255 & 0.0970 & 0.1635\\ 
         NARM & \underline{0.3191} & \underline{0.3468} & \underline{0.2578} & \underline{0.2654} & \underline{0.5162} & \underline{0.6256} & 0.1811 & 0.2519 & 0.3047 & 0.5037 & 0.1575 & 0.1182 & 0.0921 & 0.1645\\ 
         STAMP & 0.3143 & 0.3385 & 0.2558 & 0.2624 & 0.5018 & 0.5973 & 0.2704 & 0.3923 & 0.4781 & 0.8410 & 0.1977 & 0.1783 & 0.1381 & 0.2491\\    
         SR-GNN & 0.2971 & 0.3247 & 0.2371 & 0.2447 & 0.4905 & 0.5995 & 0.2435 & 0.3385 & 0.4222 & 0.7060 & 0.1836 &  0.1527 & 0.1226 & 0.2162\\ 
         GC-SAN & 0.3013 & 0.3264 & 0.2450 & 0.2520 & 0.4818 & 0.5809 & 0.2855 & 0.4014 & 0.5048 & 0.8556 & 0.2020 & 0.1774 & 0.1308 & 0.2376\\
         GCE-GNN & $\mathbf{0.3458}^{*}$ & $\mathbf{0.3723}^{*}$ & $\mathbf{0.2876}^{*}$ & $\mathbf{0.2950}^{*}$ & $\mathbf{0.5324}^{*}$ & $\mathbf{0.6373}^{*}$ & 0.1124 & 0.1623 & 0.1825 & 0.3096 & 0.1328 & 0.0892 & 0.0627 & 0.1145\\\hline
         MCPRN & 0.2321 & 0.2610 & 0.1858 & 0.1938 & 0.3829 & 0.4972 & 0.2671 & 0.3394 & 0.4651 & 0.7106 & 0.1935 & 0.1556 & 0.1100 & 0.1867  \\
         NARM+MMR & 0.2321 & 0.2454 & 0.1895 & 0.1932 & 0.3684 & 0.4204 & 0.8823 & 1.1310 & 1.7506 & 2.9703 & 0.5077 & 0.6223 & $\mathbf{0.2979}$ & $\mathbf{0.4260}$ \\ 
         IDSR($\lambda=0.2$) & 0.1677 & 0.1973 & 0.1435 & 0.1517 & 0.2521 & 0.3686 & $\mathbf{1.3108}^{*}$ & $\mathbf{1.2460}$ & $\mathbf{2.8540}^{*}$ & $\mathbf{3.3085}^{*}$ & $\mathbf{0.8127}^{*}$ & $\mathbf{0.6760}^{*}$ & \underline{0.2829} & \underline{0.4025}  \\
         IDSR($\lambda=0.5$) &0.2230 & 0.2547 & 0.1806 & 0.1893 & 0.3633 & 0.4886 & 0.7996 & 0.7531 & 1.5522 & 1.7468 & 0.4433 & 0.3401 & 0.2600 & 0.3607 \\
         IDSR($\lambda=0.8$) & 0.2681 & 0.2958 & 0.2140 & 0.2217 & 0.4438 & 0.5532 & 0.4105 & 0.4635 & 0.7464 & 1.0110 & 0.2593 & 0.2090 & 0.1814 &  0.2688\\
         \bottomrule
    \end{tabular}

\vspace{2\baselineskip}
\caption{Model Performance on Retailrocket. $*$ denotes the best model significantly outperforms the runner-up using a paired t-test (p-value $\leq 0.05$).}\label{tab:retail}
\centering
\begin{tabular}{c|cc|cc|cc|cc|cc|cc|cc}
    \toprule
     Model & \multicolumn{2}{c}{NDCG} & \multicolumn{2}{c}{MRR} & \multicolumn{2}{c|}{HR} & \multicolumn{2}{c}{ILD} & \multicolumn{2}{c}{Entropy} & \multicolumn{2}{c|}{DS} & \multicolumn{2}{c}{F-score}\\\hline
     & @10 & @20& @10 & @20& @10 & @20& @10 & @20& @10 & @20& @10 & @20& @10 & @20\\\midrule
     POP & 0.0047 & 0.0052 & 0.0026 & 0.0027 & 0.0121 & 0.0141 & \underline{1.2571} & $\mathbf{1.3334}$ & 2.4464 & \underline{3.1899} & 0.6000 & \underline{0.6667} & 0.0135 & 0.0161\\ 
     S-POP &0.3078 & 0.3086 & 0.2912 & 0.2915 & 0.3581 & 0.3612 & 1.2164 & \underline{1.2664} & 2.4831 & 3.0547 & 0.6463 & 0.6545 & $\mathbf{0.3995}^{*}$ & 0.4103\\ 
     Item-KNN & 0.1522 & 0.1598 & 0.1236 & 0.1258 & 0.2440 & 0.2729 & 0.6498 & 0.7460 & 1.2110 & 1.5946 & 0.3481 & 0.3427 & 0.1422 & 0.1871\\ 
     BPR-MF & 0.1195 & 0.1309 & 0.1000 & 0.1031 & 0.1827 & 0.2278 & 0.7872 & 0.8395 & 1.4468 & 1.8447 & 0.3896 & 0.3094 & 0.1323 & 0.1811\\ 
     FPMC & 0.0787 & 0.0918 & 0.0610 &  0.0645 & 0.1368 & 0.1885 & 0.7473 & 0.8068 & 1.3664 & 1.7411 & 0.3739 & 0.2912 & 0.0871 & 0.1361\\\hline
     GRU4Rec & 0.2380 & 0.2511 & 0.2039 & 0.2075 & 0.3467 & 0.3984 & 1.0513 & 1.1498 & 2.2385 & 3.1506 & 0.6303 & 0.6348 & 0.3075 & 0.3984\\ 
     NARM & \underline{0.3582} & \underline{0.3765} & \underline{0.3105} & \underline{0.3156} & \underline{0.5102} & \underline{0.5826} & 0.4736 & 0.5723 & 0.8401 & 1.2350 & 0.2646 & 0.2269 & 0.2451 &  0.3432\\ 
     STAMP & 0.3451 & 0.3619 & 0.3012 & 0.3058 & 0.4847 & 0.5510 & 0.5192 & 0.6375 & 0.9529 & 1.4289 & 0.2936 & 0.2641 & 0.2504 &  0.3554\\    
     SR-GNN &0.3291 & 0.3489 & 0.2829 & 0.2883 & 0.4769 & 0.5549 & 0.4811 & 0.5874 & 0.8595 & 1.2737 & 0.2713 & 0.2370 & 0.2390 & 0.3423\\ 
     GC-SAN & 0.3330 & 0.3491 & 0.2909 & 0.2954 & 0.4669 & 0.5308 & 0.5457 & 0.6660 & 1.0001 & 1.4960 & 0.3045 & 0.2756 & 0.2449 & 0.3489 \\
     GCE-GNN & $\mathbf{0.3834}^{*}$ & $\mathbf{0.4024}^{*}$ & $\mathbf{0.3355}^{*}$ & $\mathbf{0.3408}^{*}$ & $\mathbf{0.5359}^{*}$ & $\mathbf{0.6110}^{*}$ & 0.3631 & 0.4439 & 0.6122 & 0.8992 & 0.2120 & 0.1697 & 0.2121 &  0.2996\\\hline
     MCPRN & 0.2276 & 0.2409 & 0.2011 & 0.2047 & 0.3128 & 0.3653 & 0.7383 & 0.8098 & 1.4441 & 1.9635 & 0.4118 & 0.3677 & 0.2154 & 0.2766 \\
     NARM+MMR & 0.2901 & 0.2989 & 0.2551 & 0.2576 & 0.4005 & 0.4348 & 1.0292 & 1.2051 & 2.1240 & 3.2976 & 0.5993 & 0.6896 & \underline{0.3750} & \underline{0.4646}\\ 
     IDSR($\lambda=0.2$) &0.2211 & 0.2459 & 0.1983 & 0.2051 & 0.3000 & 0.3979 & $\mathbf{1.2962}$ & 1.2267 & $\mathbf{2.9170}$ & $\mathbf{3.4360}$ & $\mathbf{0.8332}$ & $\mathbf{0.7091}$ & 0.3363 & 0.4351 \\
     IDSR($\lambda=0.5$) &0.2494 & 0.2787 & 0.2254 & 0.2334 & 0.3292 & 0.4452 & 1.2677 & 1.1668 & \underline{2.7634} & 3.1378 & \underline{0.7695} & 0.6274 & 0.3652 & $\mathbf{0.4752}$\\
     IDSR($\lambda=0.8$) & 0.2192 & 0.2360 & 0.1870 & 0.1916 & 0.3238 & 0.3902 & 0.8265 & 0.8628 & 1.7779 & 2.2918 & 0.5005 & 0.4340 & 0.2430 & 0.3159\\
     \bottomrule
\end{tabular}
\end{sidewaystable}

\setlength\rotFPtop{0pt plus 1fil} 
\begin{sidewaystable}[htbp]
\footnotesize
\caption{Model Performance on Tmall. $*$ denotes the best model significantly outperforms the runner-up using a paired t-test (p-value $\leq 0.05$).}\label{tab:tmall}
\centering
    \begin{tabular}{c|cc|cc|cc|cc|cc|cc|cc}
        \toprule
         Model & \multicolumn{2}{c}{NDCG} & \multicolumn{2}{c}{MRR} & \multicolumn{2}{c|}{HR} & \multicolumn{2}{c}{ILD} & \multicolumn{2}{c}{Entropy} & \multicolumn{2}{c|}{DS} & \multicolumn{2}{c}{F-score}\\\hline
         & @10 & @20& @10 & @20& @10 & @20& @10 & @20& @10 & @20& @10 & @20& @10 & @20\\\midrule
         POP & 0.0021 & 0.0025 & 0.0011 & 0.0012 & 0.0055 & 0.0072 & 1.3199 & 1.3199 & 2.7219 & 3.0566 & 0.7000 & 0.6000 & 0.0063 & 0.0082\\
         S-POP & 0.0006 & 0.0012 & 0.0003 & 0.0004 & 0.0018 & 0.0039 & 1.2447 & 1.2763 & 2.5332 & 2.9665 & 0.6660 & 0.6075 & 0.0021 & 0.0044\\
         Item-KNN & $\mathbf{0.0321}^{*}$ & $\mathbf{0.0349}$ & $\mathbf{0.0251}^{*}$ & $\mathbf{0.0259}^{*}$ & \underline{0.0551} & \underline{0.0655} & 0.8888 & 0.9593 & 1.6790 & 2.0452 & 0.4546 & 0.4219 & \underline{0.0442} &  0.0573\\ 
         BPR-MF & 0.0096 & 0.0119 & 0.0069 & 0.0075 & 0.0186 & 0.0279 & 0.9963 & 1.0350 & 1.8716 & 2.3219 & 0.4852 & 0.3805 & 0.0168 &  0.0259\\ 
         FPMC & 0.0034 & 0.0044 & 0.0025 & 0.0028 & 0.0066 & 0.0108 & 1.0539 & 1.0969 & 1.9524 & 2.4328 & 0.4921 & 0.3883 & 0.0061 & 0.0105 \\\hline
         GRU4Rec & 0.0039 & 0.0046 & 0.0030 & 0.0032 & 0.0068 & 0.0097 & \underline{1.3415} & $\mathbf{1.3517}$ & \underline{2.9584} & $\mathbf{3.7410}$ & \underline{0.8407} & $\mathbf{0.7603}$ & 0.0075 & 0.0109\\ 
         NARM & 0.0244 & 0.0306 & 0.0174 & 0.0191 & 0.0476 & 0.0720 & 0.9453 & 1.0085 & 1.7760 & 2.2625 & 0.4689 & 0.3778 & 0.0386 &  0.0642\\ 
         STAMP & 0.0171 & 0.0215 & 0.0121 & 0.0133 & 0.0336 & 0.0511 & 1.0449 & 1.0959 & 2.0375 & 2.5806 & 0.5428 & 0.4494 & 0.0292 &  0.0481\\    
         SR-GNN & 0.0223 & 0.0274 & 0.0164 & 0.0178 & 0.0417 & 0.0619 & 1.0090 & 1.0661 & 1.9468 & 2.4823 & 0.5190 & 0.4308 & 0.0355 &  0.0574\\ 
         GC-SAN &0.0261 & 0.0324 & 0.0189 & 0.0206 & 0.0499 & 0.0747 & 0.9568 & 1.0202 & 1.8019 & 2.3003 & 0.4754 & 0.3861 & 0.0408 & \underline{0.0674} \\
         GCE-GNN & \underline{0.0282} & \underline{0.0355} & \underline{0.0187} & \underline{0.0207} & $\mathbf{0.0594}$ & $\mathbf{0.0886}^{*}$ & 0.8571 & 0.9326 & 1.5691 & 2.0340 & 0.4161 & 0.3345 & $\mathbf{0.0443}$ & $\mathbf{0.0744}$ \\\hline
         MCPRN & 0.0110 & 0.0142 & 0.0075 & 0.0084 & 0.0225 & 0.0354 & 1.0661 & 1.1042 & 2.1139 & 2.6437 & 0.5686 & 0.4679 & 0.0193 & 0.0326 \\
         NARM+MMR & 0.0235 & 0.0284 & 0.0175 & 0.0188 & 0.0432 & 0.0629 & 1.1724 & 1.2204 & 2.4576 & 3.2120 & 0.6893 & 0.6454 & 0.0409 & 0.0641 \\ 
         IDSR($\lambda=0.2$) &0.0040 & 0.0051 & 0.0036 & 0.0039 & 0.0054 & 0.0098 & 1.3274 & \underline{1.3468} & 2.8553 & 3.6406 & 0.7912 & 0.7123 & 0.0061 & 0.0112\\
         IDSR($\lambda=0.5$) &0.0055 & 0.0074 & 0.0045 & 0.0050 & 0.0089 & 0.0165 & $\mathbf{1.3483}$ & 1.3395 & $\mathbf{2.9744}$ & \underline{3.6509} & $\mathbf{0.8451}$ & \underline{0.7335} & 0.0100 &  0.0186\\
         IDSR($\lambda=0.8$) &0.0083 & 0.0114 & 0.0054 & 0.0063 & 0.0179 & 0.0303 & 1.3175 & 1.2969 & 2.8725 & 3.4530 & 0.8108 & 0.6773 & 0.0192 & 0.0327 \\
         \bottomrule
    \end{tabular}

\vspace{2\baselineskip}
\caption{Model Performance on Nowplaying. $*$ denotes the best model significantly outperforms the runner-up using a paired t-test (p-value $\leq 0.05$).}\label{tab:nowplaying}
\centering
    \begin{tabular}{c|cc|cc|cc|cc|cc|cc|cc}
        \toprule
         Model & \multicolumn{2}{c}{NDCG} & \multicolumn{2}{c}{MRR} & \multicolumn{2}{c|}{HR} & \multicolumn{2}{c}{ILD} & \multicolumn{2}{c}{Entropy} & \multicolumn{2}{c|}{DS} & \multicolumn{2}{c}{F-score}\\\hline
         & @10 & @20& @10 & @20& @10 & @20& @10 & @20& @10 & @20& @10 & @20& @10 & @20\\\midrule
         POP &0.0111 & 0.0111 & 0.0064 & 0.0064 & 0.0266 & 0.0266 & $\mathbf{1.4142}$ & $\mathbf{1.4142}^{*}$ & $\mathbf{3.3219}$ & 3.9069 & $\mathbf{1.0000}$ & $\mathbf{1.0000}^{*}$ & 0.0311 & 0.0311\\
         S-POP & 0.1051 & 0.1067 & 0.0860 & 0.0864 & 0.1683 & 0.1744 & 1.2504 & 1.3107 & 2.8207 & 3.4678 & 0.8354 & 0.8609 & 0.1517 & 0.1877\\
         Item-KNN & 0.1118 & 0.1213 & 0.0848 & 0.0876 & \underline{0.2003} & 0.2364 & 0.8879 & 0.9518 & 1.9073 & 2.3448 & 0.5716 & 0.5624 & 0.1111 & 0.1535\\ 
         BPR-MF & 0.0392 & 0.0460 & 0.0322 & 0.0341 & 0.0622 & 0.0893 & 1.3080 & 1.3298 & 2.8665 & 3.6625 & 0.8132 & 0.7427 & 0.0673 & 0.1001\\ 
         FPMC & 0.0298 & 0.0339 & 0.0250 & 0.0261 & 0.0458 & 0.0621 & 1.3176 & 1.3355 & 2.8701 & 3.6481 & 0.8081 & 0.7311 & 0.0506 &  0.0701\\\hline
         GRU4Rec & 0.0249 & 0.0275 & 0.0225 & 0.0233 & 0.0326 & 0.0432 & 1.3820 & \underline{1.3948} & 3.1797 & $\mathbf{4.1663}$ & 0.9422 & \underline{0.9402} & 0.0370 &  0.0499\\ 
         NARM & 0.1162 & \underline{0.1320} & 0.0964 & 0.1007 & 0.1810 & \underline{0.2438} & 1.1584 & 1.2095 & 2.4929 & 3.2648 & 0.7132 & 0.6638 & 0.1608 &  0.2376\\ 
         STAMP & 0.1170 & 0.1300 & 0.0982 & 0.1017 & 0.1785 & 0.2300 & 1.2241 & 1.2686 & 2.6837 & 3.5309 & 0.7749 & 0.7421 & \underline{0.1703} &  0.2367\\    
         SR-GNN & 0.1160 & 0.1311 & 0.0960 & 0.1001 & 0.1818 & 0.2415 & 1.1985 & 1.2446 & 2.6096 & 3.4242 & 0.7512 & 0.7112 & 0.1686 & \underline{0.2432} \\ 
         GC-SAN & \underline{0.1182} & 0.1307 & \underline{0.1016} & \underline{0.1050} & 0.1725 & 0.2220 & 1.2160 & 1.2611 & 2.6557 & 3.4900 & 0.7651 & 0.7292 & 0.1625 & 0.2268 \\
         GCE-GNN & $\mathbf{0.1293}^{*}$ & $\mathbf{0.1480}^{*}$ & $\mathbf{0.1038}^{*}$ & $\mathbf{0.1090}^{*}$ & $\mathbf{0.2130}^{*}$ & $\mathbf{0.2870}^{*}$ & 1.0471 & 1.1264 & 2.2147 & 2.9912 & 0.6367 & 0.6065 & \underline{0.1707} & $\mathbf{0.2616}^{*}$ \\\hline
         MCPRN & 0.0521 & 0.0640 & 0.0401 & 0.0432 & 0.0922 & 0.1394 & 1.2330 & 1.2684 & 2.7035 & 3.5214 & 0.7790 & 0.7368 & 0.0900 & 0.1428 \\
         NARM+MMR & 0.1079 & 0.1201 & 0.0906 & 0.0939 & 0.1646 & 0.2128 & 1.2984 & 1.3499 & 2.9283 & 3.9441 & 0.8614 & 0.8758 & $\mathbf{0.1713}$ & 0.2382 \\ 
        IDSR($\lambda=0.2$) &0.0481 & 0.0586 & 0.0447 & 0.0476 & 0.0602 & 0.1018 & \underline{1.4000} & 1.3872 & \underline{3.2393} & \underline{4.1021} & \underline{0.9608} & 0.9131 & 0.0700 & 0.1167\\
        IDSR($\lambda=0.5$) &0.0548 & 0.0693 & 0.0493 & 0.0532 & 0.0738 & 0.1318 & 1.3635 & 1.3533 & 3.0846 & 3.9018 & 0.9001 & 0.8487 & 0.0830 & 0.1467\\
        IDSR($\lambda=0.8$) & 0.0628 & 0.0769 & 0.0521 & 0.0559 & 0.0993 & 0.1552 & 1.3670 & 1.3358 & 3.1101 & 3.8045 & 0.9141 & 0.8186 & 0.1124 & 0.1719\\
         \bottomrule
    \end{tabular}
\end{sidewaystable}

\subsubsection{Performance on Recommendation Accuracy}
As shown in Tables~\ref{tab:digi}-\ref{tab:nowplaying}, the performance of different approaches on recommendation accuracy is measured via NDCG@$N$, MRR@$N$, and HR@$N$ ($N=\{10, 20\}$). From the results in Tables~\ref{tab:bordacount}-\ref{tab:nowplaying}, several interesting observations are noted. 

(1) Regarding recommendation accuracy, deep neural methods generally outperform traditional non-neural methods, except that Item-KNN performs the best on Tmall. The deep diversified methods (i.e., MCPRN, {NARM+MMR}, and IDSR with $\lambda=\{0.2,0.5,0.8\}$ respectively) perform worse than the accuracy-oriented deep methods in most cases, but better than traditional ones. 
(2) Among traditional non-neural methods, the performance of S-POP and Item-KNN could be within the same order of magnitudes with that of deep neural methods, except for S-POP on Tmall. However, MF-based method like 
BPR-MF, performing quite well in traditional RSs, gains relatively worse accuracy in SBRSs. 
(3) In regard to the six (accuracy-oriented) deep neural methods, GCE-GNN, which further considers global graph instead of only session graph like other GNN-based models (GC-SAN and SR-GNN), achieves the best performance in all scenarios. NARM using vanilla attention ranks second in most scenarios (except on Tmall), which, however, has lower computational complexity than GCE-GNN. 
(4) Of the diversified methods, IDSR (with some $\lambda$) can defeat MCPRN in most scenarios, except on Tmall (see Tables~\ref{tab:digi}-\ref{tab:nowplaying}); whereas, the overall accuracy performance of MCPRN exceeds that of IDSR (see Table \ref{tab:bordacount}). Besides, the performance of IDSR w.r.t. accuracy metrics generally improves with the increase of $\lambda$ value (except on Retailrocket), conforming to the intuition of trade-off hyper-parameter in model design.
{
In addition, in terms of accuracy,
NARM+MMR is the top-ranked method among deep diversified methods, but it drops down four positions compared to NARM. This is because NARM focuses primarily on accuracy during its learning process, while MMR is a re-ranking technique that emphasizes diversity without learning. Therefore, even a slight increase in the $\lambda$ multiplier used in MMR can result in a significant reduction in the accuracy of NARM+MMR. Thus, with an optimal and small $\lambda$ value ($0.005$), NARM+MMR avoids the large decrease in accuracy resulting from diversity-promoting re-ranking, and thus achieves the best comprehensive performance (F-score). In contrast to other diversified models, NARM+MMR still performs competitively in terms of accuracy.}

\subsubsection{Performance on Recommendation Diversity}
As shown in Tables~\ref{tab:digi}-\ref{tab:nowplaying}, the performance of different approaches on recommendation diversity is measured via ILD@$N$, Entropy@$N$, and DS@$N$ ($N=\{10, 20\}$). Based on Tables~\ref{tab:bordacount}-\ref{tab:nowplaying}, we gain three observations.

(1) Regarding recommendation diversity, the diversified method, IDSR ($\lambda=0.2$) performs the best as it ranks first on Diginetica and Retailrocket or second on Nowplaying, except that IDSR ($\lambda=0.5$) performs better on Tmall. Besides, non-neural methods (POP and S-POP), and accuracy-oriented deep method (GRU4Rec), though not being specifically designed for diversity purpose, obtain a relatively better performance than other baselines, which is comparable to IDSR. (2) The accuracy-oriented deep methods (except GRU4Rec), especially GCE-GNN and NARM, usually perform far behind other baselines w.r.t. recommendation diversity.
(3) The deep diversified methods beat the accuracy-oriented deep methods. MCPRN, though being worse than IDSR, outperforms other accuracy-oriented SBRSs in most cases. {
Furthermore, the use of MMR with NARM can improve diversity performance in comparison to NARM alone. This is because MMR reranks candidate items in terms of a diversification score. However, the performance of NARM+MMR is suboptimal compared to other diversified methods due to the low $\lambda$ value, which is intended to balance accuracy and diversity while maximizing the F-score and preventing a significant decrease in accuracy in exchange for increased diversity.}
Moreover, the performance of IDSR on recommendation diversity decreases with the increase of $\lambda$ (except on Tmall dataset), which is also consistent with the trade-off hyper-parameter's intuition.

\subsubsection{Comprehensive Performance}
By following the idea of \cite{hu2017diversifying}, in order to more thoroughly and fairly assess session-based algorithms from both accuracy and diversity perspectives, we have compared them in terms of F-score (see Equation \ref{eq:F-score}). It should be noted that we can compute the Harmonic mean of any two combinations where one comes from accuracy group (i.e., HR, MRR, and NDCG) and the other from diversity group (i.e., ILD, Entropy, and DS). As can be seen from Tables~\ref{tab:digi}-\ref{tab:nowplaying} and Figure \ref{fig:corr_by_da},
the metrics within the same group are positively correlated with each other. Thus, for evaluating the comprehensive performance, we particularly select the most popular one from each group respectively, i.e., HR and ILD, to calculate the F-score. As shown in Tables~\ref{tab:bordacount}-\ref{tab:nowplaying}, we have the following observations regarding the F-score.

(1) Among the non-neural methods, POP, although it performs quite well on diversity, ranks the last in most scenarios w.r.t. the comprehensive performance except on Tmall. Nevertheless, S-POP, which far exceeds POP on accuracy due to its particular design for session situation, ranks the first of the five traditional methods except on Tmall.
(2) Generally speaking, deep neural methods (including the diversified ones) obtain better comprehensive performance than non-neural traditional methods, where {NARM+MMR} is the winner on Diginetica and Retailrocket, and GCE-GNN ranks first on Tmall and Nowplaying.
{Applying MMR to NARM with the optimal $\lambda$ value (highest F-score) results in a more balanced performance compared to using NARM alone. 
Table \ref{tab:bordacount} demonstrates that NARM+MMR ranks highest overall, followed closely by IDSR ($\lambda=0.5$).}
(3) For the diversified methods, IDSR (with some $\lambda$) and {NARM+MMR} consistently outperform MCPRN. Besides, MCPRN underperforms other deep neural methods.

To conclude, from the above results, we can see that, (1) accuracy-oriented approaches (e.g., GC-SAN, STAMP), although not particularly designed for diversity purpose, can achieve a satisfying balance between accuracy and diversity, and thus gain better comprehensive performance. And, the deep diversified method, although emphasizing more on diversity, can outperform traditional non-neural methods on recommendation accuracy.
{NARM+MMR and IDSR show strong overall performance by achieving satisfactory results on both evaluation metrics}; and (2) the performance of different approaches regarding accuracy, diversity and  F-score varies across the four datasets, which will be detailed in the following subsections.

\subsubsection{In-depth Analysis for Varied Performance across Approaches (Case Study)}\label{sec:in-depth}

\begin{figure}
\centering
    \includegraphics[scale=0.42]{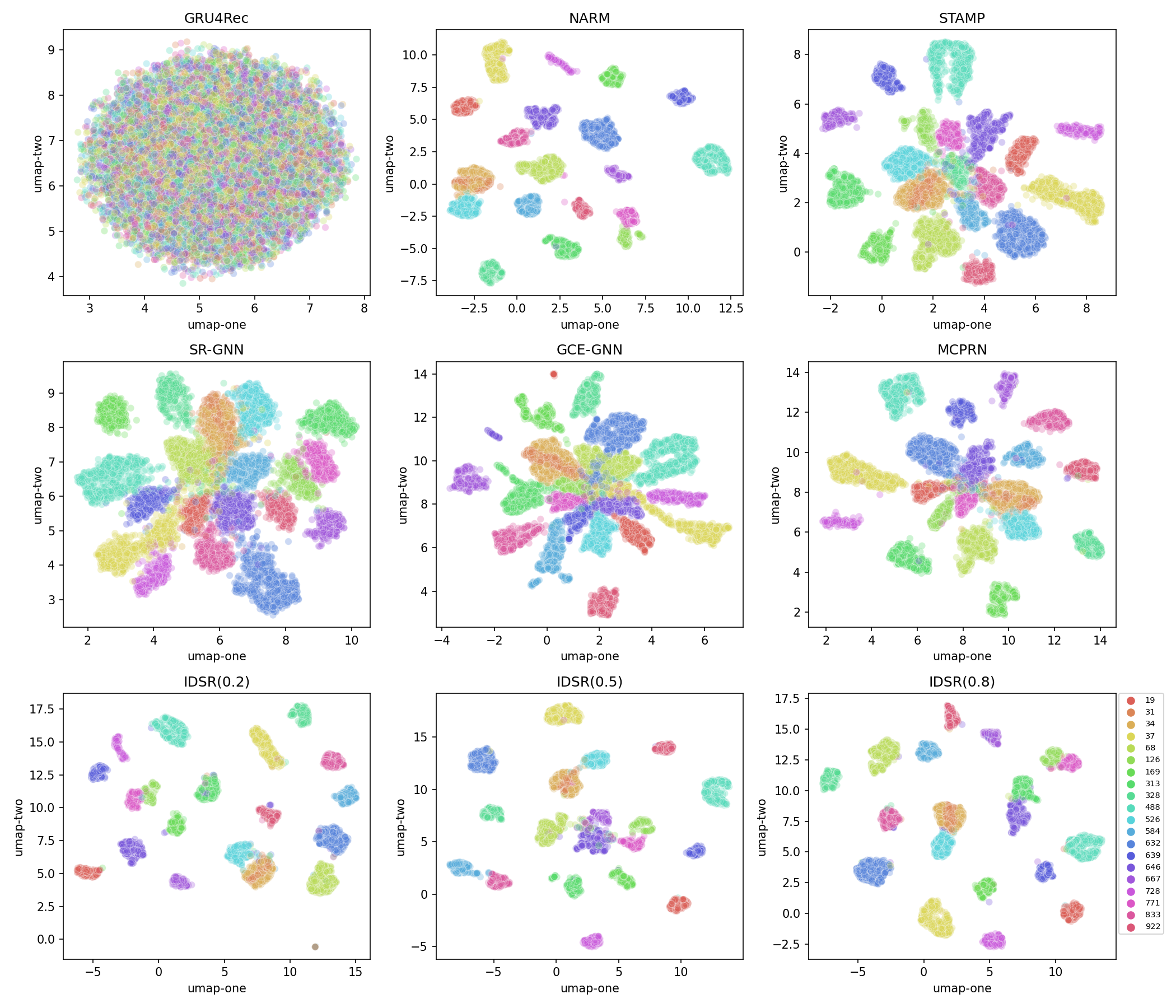}
    \caption{{Reduced Dimensional Embeddings (Using UMAP) Of Items From the Most Popular $20$ Categories.
    }}
    \label{fig:umap_digi_20}
\end{figure}

\begin{figure}
\centering
    \includegraphics[scale=0.47]{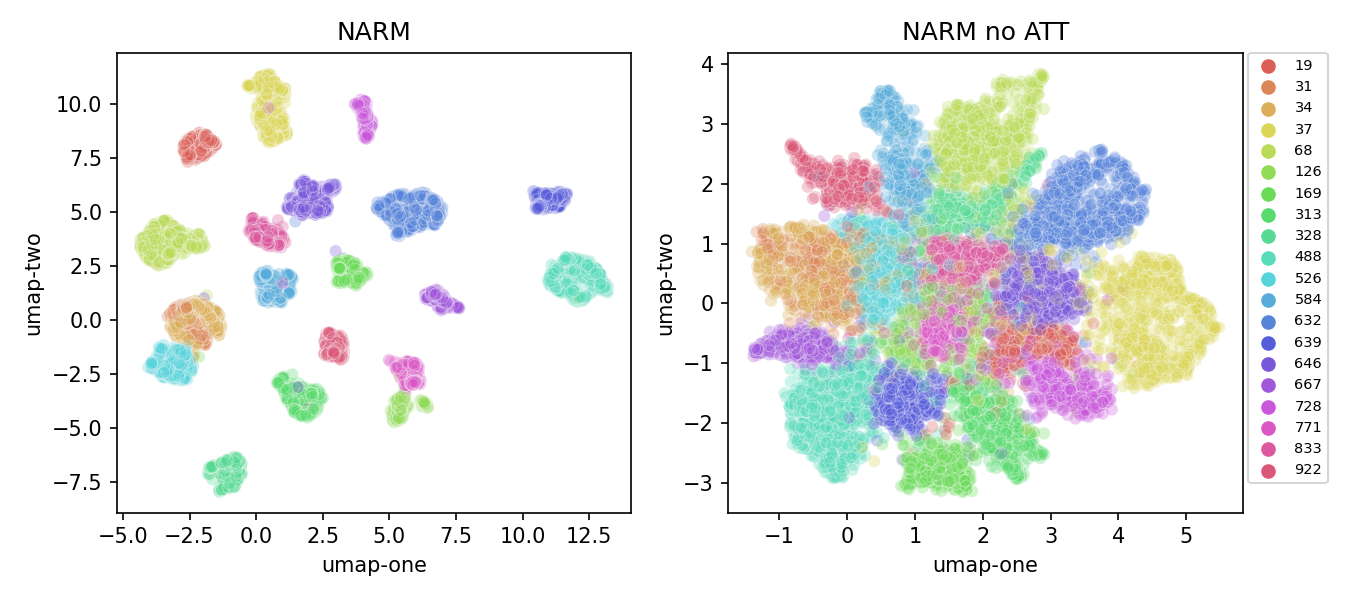}
    \caption{{The Impact of Attention Mechanisim on Reduced Dimensional Embeddings (on Diginetica)}}
    \label{fig:umap_digi_noatt_20}
\end{figure}

Here, we strive to analyze the underlying possible reasons leading to the varied performance of different approaches.
To fulfill the goal, we have investigated the item embeddings obtained by different approaches. This is mainly because the learned item embeddings are aggregated with varied model-designs (e.g., RNN, Attention mechanisms, and GNNs) to learn session representations (implying anonymous user preferences), which are then combined back with item embeddings (mostly in the form of inner product~\cite{li2017neural,WuT0WXT19,wang2020global}) to generate item ranking scores for next-item prediction.

In particular, we visualize the learned item embeddings by different approaches using Diginetica dataset as our case study (similar results can be obtained on other datasets) in Figure~\ref{fig:umap_digi_20}. For ease of presentation, we map the high-dimensional item embeddings learned by different baselines into a two-dimensional space. Particularly, we choose the dimension reduction method UMAP (Uniform Manifold Approximation and Projection)~\cite{mcinnes2018umap} which is competitive with t-SNE~\cite{van2008visualizing} for visualization quality, and arguably preserves more information of the global structure with superior run time performance. Besides, we label items of the same category with the same color. And, in Figure \ref{fig:umap_digi_20}, we only show items of the most popular $20$ categories (which refer to those categories including the largest number of items in our paper) to better view the results\footnote{Since the number of categories on these datasets is relatively large (e.g., $995$ on Diginetic), plotting and interpreting all of them with different colors are extremely challenging. Thus, we choose to plot items from the most popular $20$ categories. Similar results can be observed in terms of items from the most popular $50$ categories (see Figure~\ref{fig:umap_digi_50} in Appendix). Figure \ref{fig:umap_digi_total} in Appendix shows embeddings of all items without labeling categories with different colors.}.

Furthermore, as non-diversified accuracy-oriented deep neural methods can well balance between accuracy and diversity though not particularly designed for diversity purpose, we specifically select five representative ones to be explored: GRU4Rec (RNN-based), NARM and STAMP (Attention-based), and SR-GNN and GCE-GNN (GNN-based). {Besides, as a comparison, we also display the results of the two deep diversified methods: MCPRN and IDSR ($\lambda\in \{0.2, 0.5, 0.8\}$, respectively. And, for example, IDSR($0.2$) refers to IDSR with $\lambda=0.2$)\footnote{{We do not present that of NARM+MMR since it is built on NARM's learned item embeddings.}}. 
It is important to note that on Diginetica, MCPRN is a one-channel model (i.e., non-diversified one) as shown in Table~\ref{tab:optimalhypers}, which suggests that this is the optimal configuration for MCPRN.
In contrast, IDSR employs both a diversification score and a relevance score, making it difficult and meaningless to compare item embeddings alone without taking into account the effect of the model's unique design on accuracy and diversity.
}
The item embeddings using UMAP of the most popular $20$ for these approaches are shown in Figure~\ref{fig:umap_digi_20}, where items (dots) of a same color are from the same category.
Recall that as can be seen in Table~\ref{tab:bordacount}, in terms of accuracy metrics, the rankings of these methods are: GCE-GNN $>$ NARM $>$ STAMP $>$ SR-GNN {$>$ IDSR($0.8$)} {$>$ MCPRN} {$>$ IDSR($0.5$)} {$>$ IDSR($0.2$)} $>$ GRU4Rec, while those for diversity are: {IDSR($0.2$) $>$} GRU4Rec $>$ {IDSR($0.5$) $>$} {IDSR($0.8$) $>$} STAMP {$>$ MCPRN} $>$ SR-GNN $>$ NARM $>$ GCE-GNN.
As can be seen in Figure~\ref{fig:umap_digi_20}, towards these methods, we have the following observations.

(1) The learned item embeddings of different categories by GRU4Rec are mixed and inseparable, accounting for the highest diversity and lowest accuracy, since ranking score of an item is the inner product of the item embedding and the corresponding session representation. {
Additionally, it has been observed that MCPRN's learned item embeddings exhibit more distinct boundaries among various categories in comparison to those of GRU4Rec. This can be attributed to MCPRN's utilization of a specialized recurrent unit, the purpose-specific recurrent unit (PSRU), which is a modified variant of GRU. The improved separation of categories achieved through clearer boundaries facilitates the more accurate recommendation as compared to GRU4Rec.}
(2) Regarding NARM, we can see that the learned item embeddings of different categories are clearly distinguishable from each other. The possible reason is, with attention mechanism, main purpose is captured, which {can} cause the embeddings of items in a session to be greatly differed if they do not relate to the main-purpose.
{To validate the impact of the attention mechanism, taking the representative attention-based method NARM as an example, 
we compared the learned embedding distribution between the NARM model with and without the attention mechanism on the Diginetica dataset. The results, as shown in Figure~\ref{fig:umap_digi_noatt_20}, demonstrate the distinguishable capacity of the attention mechanism for different categories.}
On the other hand, it also explains why NARM improves over GRU4Rec on accuracy, but obtains much worse diversity performance. (3) STAMP obtains quite similar patterns on item embeddings, compared to NARM, but the distance between items of different categories is smaller. From the model perspective, we know that, in contrast to NARM which applies GRU and attention mechanism, STAMP combines MLP and attention mechanism where the exploited simple MLP in STAMP work worse than GRU for capturing sequential information. Therefore,
STAMP achieves a much better performance on diversity but worse performance on accuracy than NARM. 
(4) Concerning the two GNN-based approaches, both SR-GNN and GCE-GNN can distinguish the items of different categories, that is, item embeddings of different categories vary with each other, as viewed in Figure \ref{fig:umap_digi_20}. On the other hand, compared to attention-based methods, they achieve much smaller distance between learned item embeddings of different categories.
This is mainly because that GNN is capable of capturing more complex relationships among items in a session and across different sessions, leading to the reduced distance between items, which explains the improved diversity of SR-GNN compared to NARM though they both consider attention mechanism and item relationship in a session. Furthermore, GCE-GNN further incorporates item relationships across sessions using the global graph, which can facilitate item learning (increased distance between items of different categories) and user preferences learning, and thus obtain better accuracy but worse diversity.
{(5) Regarding IDSR, we have observed that as we increase the weight of the relevance score and decrease the weight of the diversification score by changing $\lambda$ from $0.2$ to $0.8$ in IDSR, the number of instances in which the blended embeddings of items from different categories decrease. In other words, there is a decrease in the number of cases where there is a clear distinction between blended embeddings of items from different categories.}

\begin{figure*}
\captionsetup[subfloat]{farskip=2pt,captionskip=1pt}
    \centering
     \subfloat[Uniformity of NARM]{
         \label{fig:narm_digi_uniformity}
         \includegraphics[width=1.\linewidth]{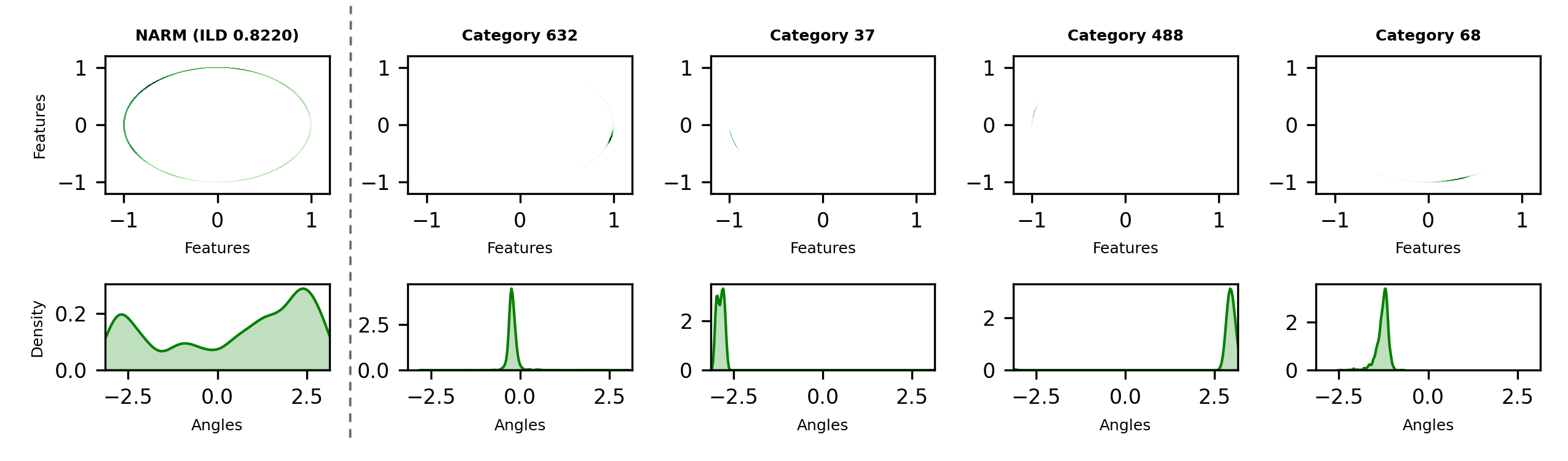}}\\
     \subfloat[Uniformity of GCE-GNN]{
         \label{fig:gce_digi_uniformity}
         \includegraphics[width=1.\linewidth]{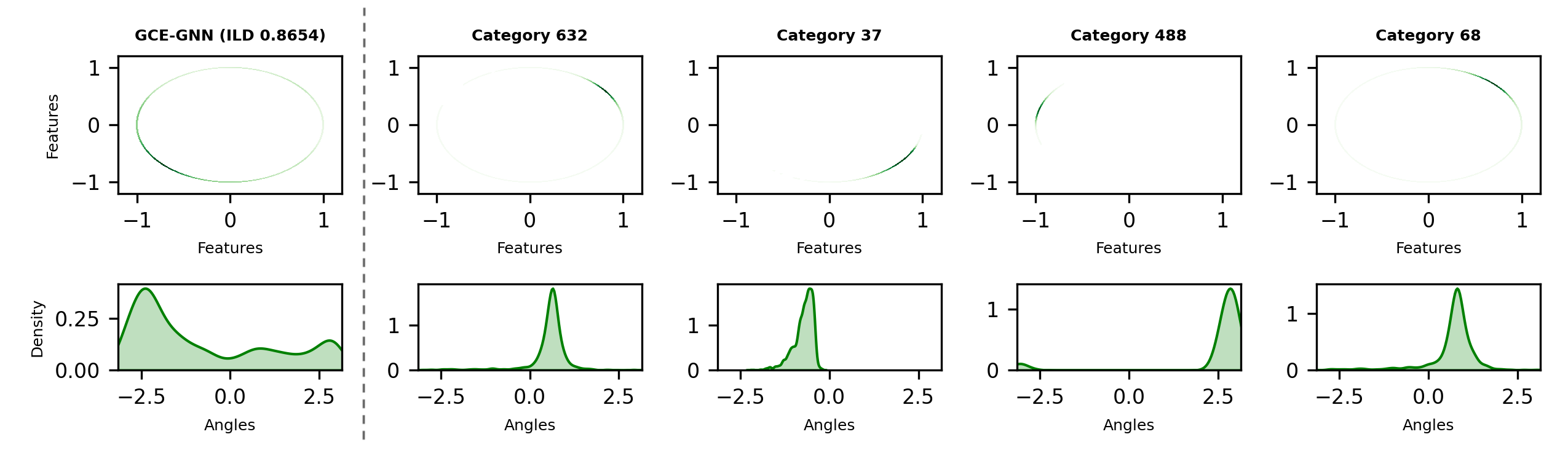}}\\
     \subfloat[Uniformity of STAMP]{
         \label{fig:stamp_digi_uniformity}
         \includegraphics[width=1.\linewidth]{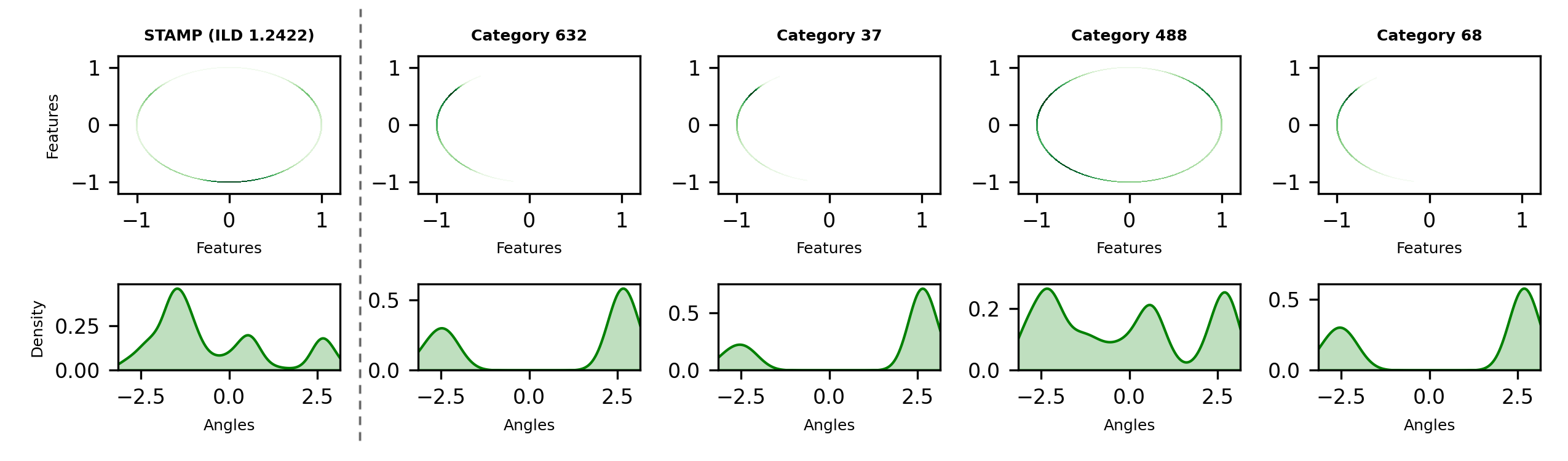}}\\
     \subfloat[Uniformity of GRU4Rec]{
         \label{fig:gru4rec_digi_uniformity}
         \includegraphics[width=1.\linewidth]{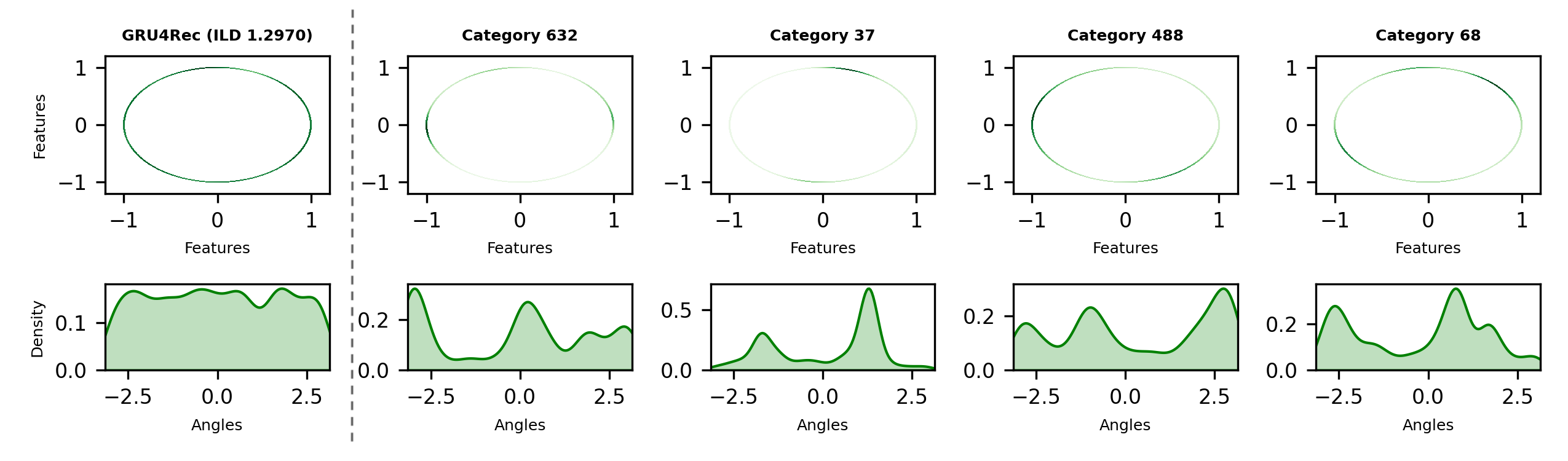}}
    \caption{
    Uniformity analysis on the distribution of item embeddings learned from the Diginetica dataset using Gaussian kernel density estimation (KDE) in $\mathbb{R}^2$ (the density of points represented by darker colors) and von Mises-Fisher (vMF) KDE on angles (i.e., $arctan2(y, x)$ for each point $(x, y) \in S^1$).
    }
    \label{fig:uniformity}
\end{figure*}

{
Furthermore, inspired by the principle of uniformity~\cite{wang2020understanding,yu2022graph}, we apply a normalization technique to the item embeddings, thus mapping them into a unit circle. Subsequently, we examine the correlation between the diversity performance and the degree of overlap observed among items belonging to different categories. Figure~\ref{fig:uniformity} depicts the uniformity of GRU4Rec and STAMP (with superior diversity) as well as NARM and GCE-GNN (with poor diversity), where the distribution of all item embeddings is shown to the left of the dotted line; while the distribution of item embeddings for each of the most popular categories is displayed to the right, in turn with category 623, 37, 488, and 68, respectively. Based on Figure~\ref{fig:uniformity}, we note that the high degree of uniformity in the distribution of all item representations may not be able to completely help explain the superior diversity performance. For instance, although NARM possesses higher uniformity compared with STAMP, it achieves worse diversify performance (ILD: 0.8220) than STAMP (ILD: 1.2422). Contrarily, the distribution of different categories regarding uniformity can provide insight into the performance of diversity. In particular, for NARM (Figure~\ref{fig:narm_digi_uniformity}) and GCE-GNN (Figure~\ref{fig:gce_digi_uniformity}) with relatively worse diversity performance, the distribution of item embeddings from the most popular categories is distinguishable, namely low degree of overlap among different categories (i.e., lower uniformity); whilst for STAMP (Figure~\ref{fig:stamp_digi_uniformity}) and GRU4Rec (Figure~\ref{fig:gru4rec_digi_uniformity}) with relatively better diversity performance, the distribution of item embeddings from the most popular categories is similar, namely high degree of overlap among different categories (higher uniformity). The possible explanation is that the high degree of overlap among various categories leads to the non-differentiable items from different categories in the final recommendation, thus enhancing the diversity of the recommendation lists, vice versa.
}

To conclude, different types of neural network structures work differently regarding recommendation accuracy and diversity: (1) RNN-based methods, although capturing sequential information in a session, can not well distinguish items (due to the involved noise), and thus cannot work well on accuracy; (2) attention mechanism can well address noisy information and distinguish item information, however, simply depending on the item sequential relationship in a session {can} overlook some information for accurate item learning; and (3) GNN-based methods can relatively well tackle the issues suffered by RNN-based and attention-based methods, and thus achieve better recommendation accuracy, but obtain less diverse item embeddings
regarding categories leading to worse performance on diversity. 
{Furthermore, the even distribution across various categories contributes to the diversity performance.}

\subsection{Accuracy-Diversity Relationship (RQ2)}\label{sec:acc-div-rel}

\begin{figure}
    \centering
    \includegraphics[scale=0.5]{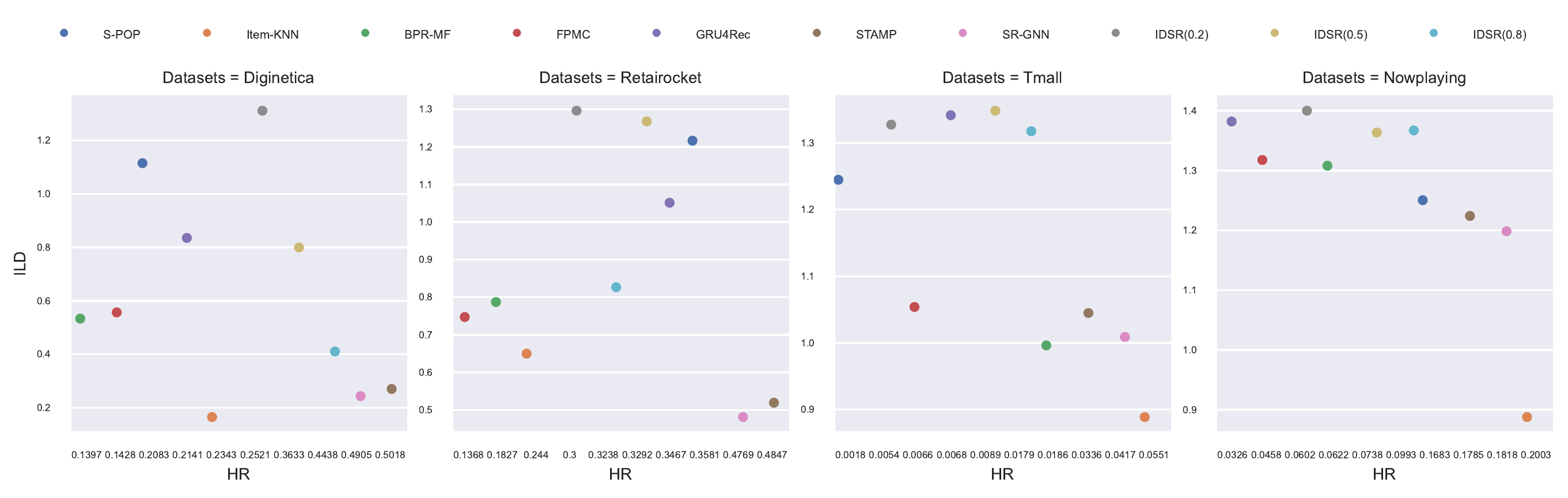}
    \caption{{The HR-ILD scatter Diagram of Some Baselines with $N=10$.}}
    \label{fig:hr_ild_scatter}
\end{figure}

Accuracy-Diversity Trade-off (Dilemma)~\cite{zhou2010solving,zheng2021dgcn} refers to that the performance improvement on diversity can only be taken place at the expense of recommendation accuracy and vice versa. 
Inspired by the intuition, most of the diversified recommendation methods further design a trade-off hyper-parameter to combine relevance score (for accuracy) and diversification score, e.g.,~\cite{carbonell1998use,chen_fast_2018,chen2020improving}.
Particularly, IDSR clearly shows such kind of dilemma on accuracy and diversity (accuracy improves and simultaneously diversity decreases when $\lambda=0.2\rightarrow 0.5\rightarrow 0.8$) as shown in Tables~\ref{tab:digi} and \ref{tab:nowplaying}. However, we can also see that, although equipped with such trade-off design, IDSR fails to prove the trade-off relationship on Retailrocket and Tmall, where a lose-lose relationship can be found between diversity and accuracy in Table~\ref{tab:retail} ($\lambda=0.5\rightarrow 0.8$), and a win-win one is present in Table~\ref{tab:tmall} ($\lambda=0.2\rightarrow 0.5$). {To facilitate the identification of win-win scenarios, we visualize certain outcomes obtained from Tables \ref{tab:digi}-\ref{tab:nowplaying} in Figure~\ref{fig:hr_ild_scatter}. This includes comparisons such as S-POP vs. (BPR-MF and FPMC) on Diginetica and Retailrocket, as well as STAMP vs. SR-GNN.}

\begin{figure}
    \centering
    \includegraphics[scale=0.45]{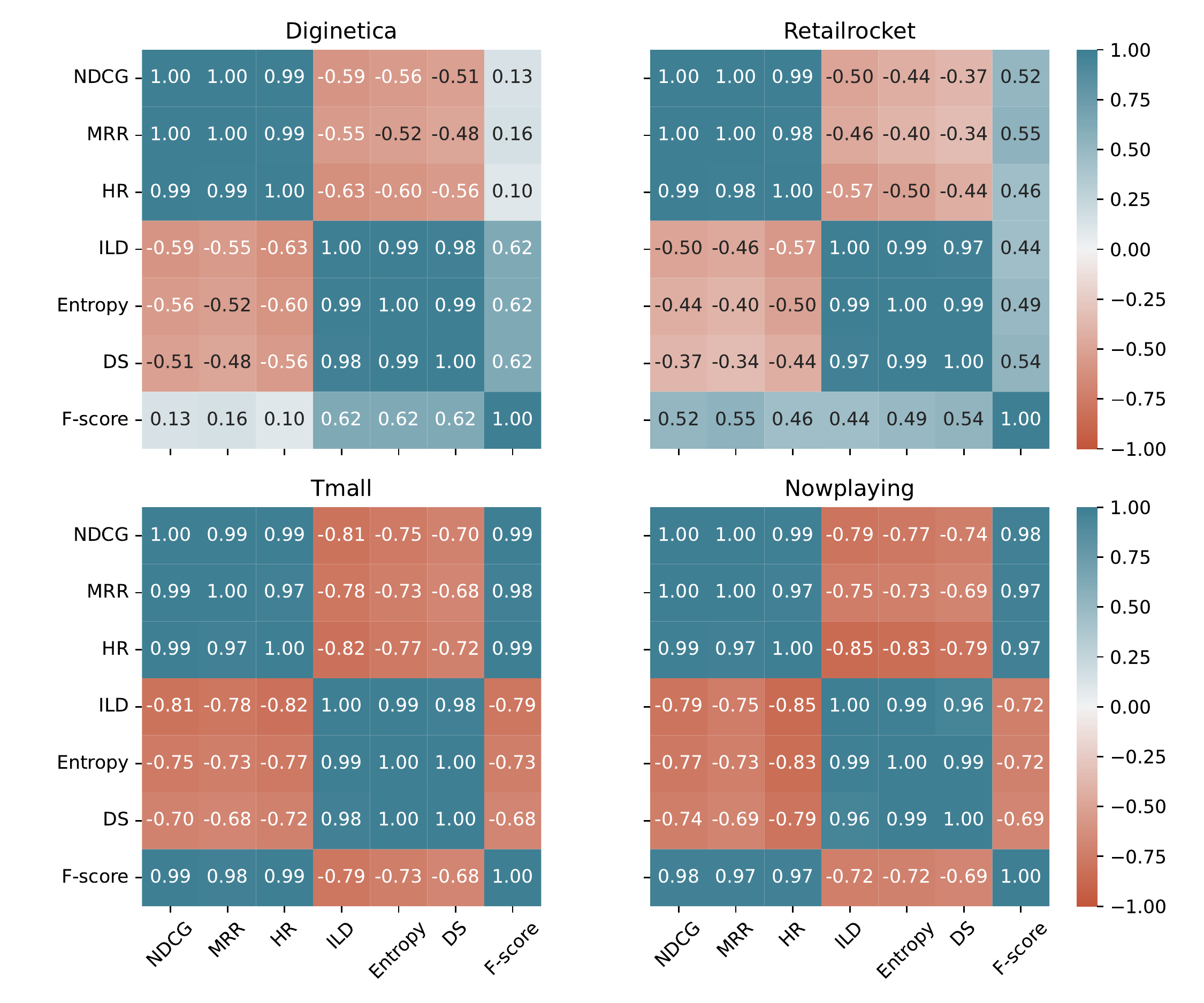}
    \caption{Pearson Correlation Coefficient of Metrics on Every Dataset. Each value is calculated given two corresponding arrays by concatenating Top-$10$ performance of baselines on the respective dataset.}
    \label{fig:corr_by_da}
\end{figure}

To further analyze, the win-win relationship can be obtained by properly mining user preferences.
On a dataset where user preferences are more diversified, it is more likely to view the same-trend for both accuracy and diversity. In this case, blindly pursuing diversity will make accuracy deteriorate. On the contrary, recommendation accuracy can be maintained or even boosted 
if personalized diversity is reasonably considered~\cite{wu2019pd}. For example, as shown in Tables~\ref{tab:digi}-\ref{tab:nowplaying}, POP and S-POP always provide well-diversified recommendations due to popularity selection. By better fusing personalized user preference regarding every session, S-POP obtains much better accuracy while only sacrificing little performance on diversity compared to POP, {thus achieving win-win compared with other traditional methods {(i.e., Item-KNN, BPR-MF, and FPMC)} 
on Diginetica and Retailrocket {shown in Figure~\ref{fig:hr_ild_scatter}}.}
{
The above satisfying performance of S-POP is due to its model design and its conformity with the unique property of Diginetica and Retailrocket. Specifically, 1) typically the session length is limited (e.g., avg.len.<5 on Diginetica and Retailrocket shown in Table~\ref{tab:datasetStatistics}), so is the number of unique items in the session. That is, besides the unique items in the session sorted by frequency, S-POP completes the Top-N (e.g., $N = \{10, 20\}$) recommendation list with the most frequent items based on global popularity which increases the diversity performance. 2) Compared to Tmall and Nowplaying with a lower repeat ratio~\cite{repeat2019} (i.e., items that appeared repeatedly within a session) at 0\% and 4\%, respectively, whereas Diginetica and Retailrocket have a larger repeat ratio at 13\% and 24\%. As such, the high repeat ratio property of Diginetica and Retailrocket enables the S-POP with the most frequent item recommendation strategy to achieve competitive accuracy.
}

{In summary, the relationship between accuracy and diversity is quite complex and mixed. Besides the “trade-off” relationship, they can be positively correlated with each other, that is, possessing a same-trend (win-win or lose-lose). Such as the special model design (like IDSR) and unique characteristics of datasets (e.g., the aforementioned win-win scenario on Diginetica and Retailrocket from S-POP) can be the reasons leading to varied observed relationship between diversity and accuracy.}
{
For instance, when using the same type of model, IDSR, with different values of $\lambda$ to adjust the importance of relevance score and diversification score, we can observe different trends across datasets with varying diversity scores. Tmall and Nowplaying have higher diversity scores than Diginetica and Retailrocket, and the win-win situations of $\lambda= 0.2 \rightarrow 0.5$ on Tmall and $\lambda= 0.5 \rightarrow 0.8$ on Nowplaying in Figure~\ref{fig:hr_ild_scatter} demonstrate the mutual benefits between accuracy and diversity.
}

Next, we seek to deeply explore the varied relationship from the inter- and intra-model views.
From the inter-model view, as observed in Tables~\ref{tab:digi}-\ref{tab:nowplaying} {and Figure~\ref{fig:hr_ild_scatter}}, besides the aforementioned same-trend cases, GRU4Rec gains better accuracy and diversity than FPMC and BPR-MF on Diginetica and Retailrocket, and so is the STAMP to SR-GNN.
The trade-off relationship occurs more commonly, especially for deep neural methods with promising accuracy performance. With accuracy as the main objective, they often show the dilemma on increasing accuracy with the decrease of diversity, e.g., GCE-GNN ranks the first place in terms of accuracy but the last one on diversity.
Besides, with regard to datasets, the same-trend relationship is more common for every model on Retailrocket. To dig more, we calculate the Pearson Correlation Coefficients among the seven adopted metrics by gathering all baselines' performance (@10) on each dataset. As presented in Figure~\ref{fig:corr_by_da}, overall, the trade-off relationship is dominating between accuracy and diversity, as the correlation between every metric in accuracy group and diversity group is negative. However, the trade-off relationship is the weakest on Retailrocket (with smallest absolute values of negative coefficients), followed by the second weakest of Diginetica, compared to the other two datasets. This explains why we can view more same-trend cases on Retailrocket and Diginetica.

\begin{figure}
\centering
    \includegraphics[scale=0.41]{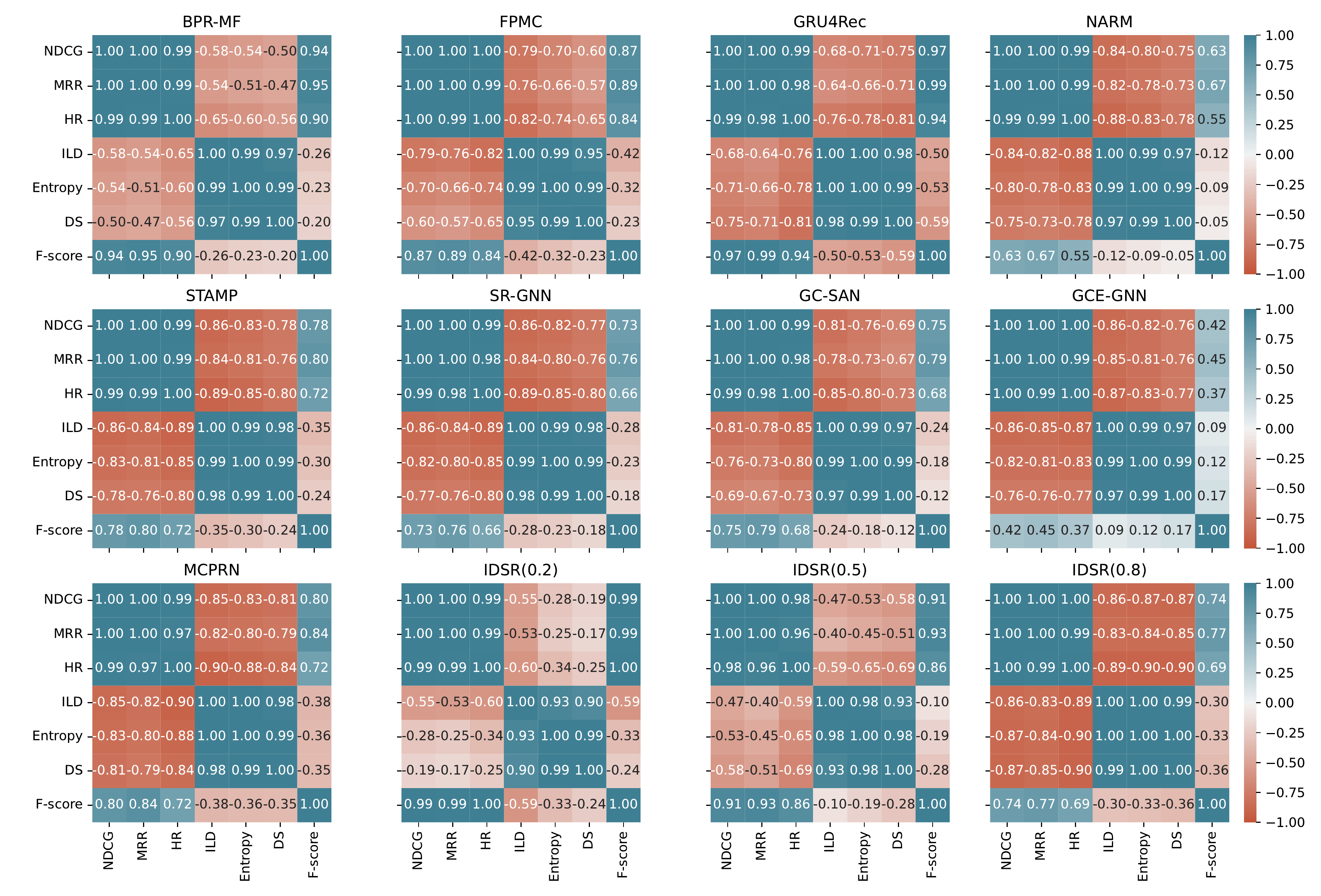}
    \caption{Pearson Correlation Coefficient of Metrics Regarding Different Baselines. Each value is calculated given two corresponding arrays by concatenating Top-$10$ performance of each baseline on all the datasets.}
    \label{fig:corr_by_md}
\end{figure}

\begin{figure}
    \centering
    \includegraphics[scale=0.52]{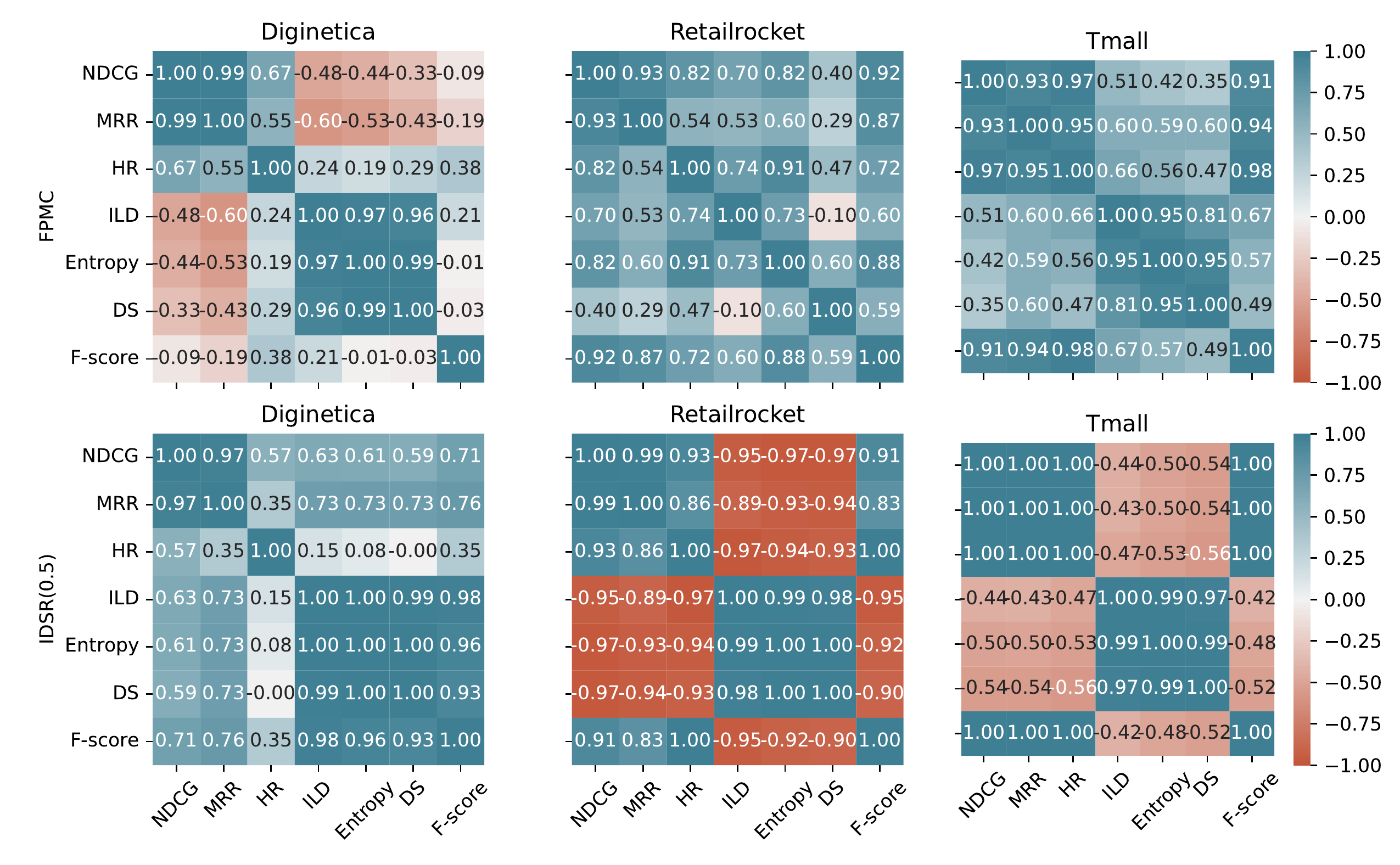}
    \caption{Pearson Correlation Coefficient of Metrics for Different Baselines on Every Dataset. Each value is calculated given two arrays by concatenating Top-10 performance (running 5 times) of each baseline on each dataset.}
    \label{fig:corr_by_md_da}
\end{figure}

{From the intra-model view, for further analysis, we also calculate the Pearson Correlation Coefficient for each learnable baseline (excluding POP, S-POP, and NARM+MMR) among the seven metrics (@10) by gathering their performance on all four datasets.}
The results are depicted in Figure~\ref{fig:corr_by_md}, where the trade-off relationship is relatively weaker on BPR-MF, FPMC, GRU4Rec and IDSR ($\lambda=\{0.2,0.5\}$) (with smaller absolute values of negative coefficients) compared to other baselines. 
{
At a more granular level, we conduct a detailed analysis of each method on every dataset (see Figure \ref{fig:corr_by_md_da}\footnote{{To save space, we display a subset of results of different baselines on every dataset. And, the whole results can be found in Figure~\ref{fig:corre_bydataandmodel_total} in Appendix.}}) using five data points, as each model was run on each dataset five times with optimal hyperparameters.}
We note that the relationship between accuracy and diversity regarding every method varies across different datasets, where the same-trend relationship (positive coefficients) also can be observed. 

To sum up, from the inter- and intra-model perspectives, it is revealed that, besides the trade-off relationship, same-trend one does exist across different datasets and methods. 

\begin{figure}
    \centering
    \includegraphics[scale=0.53]{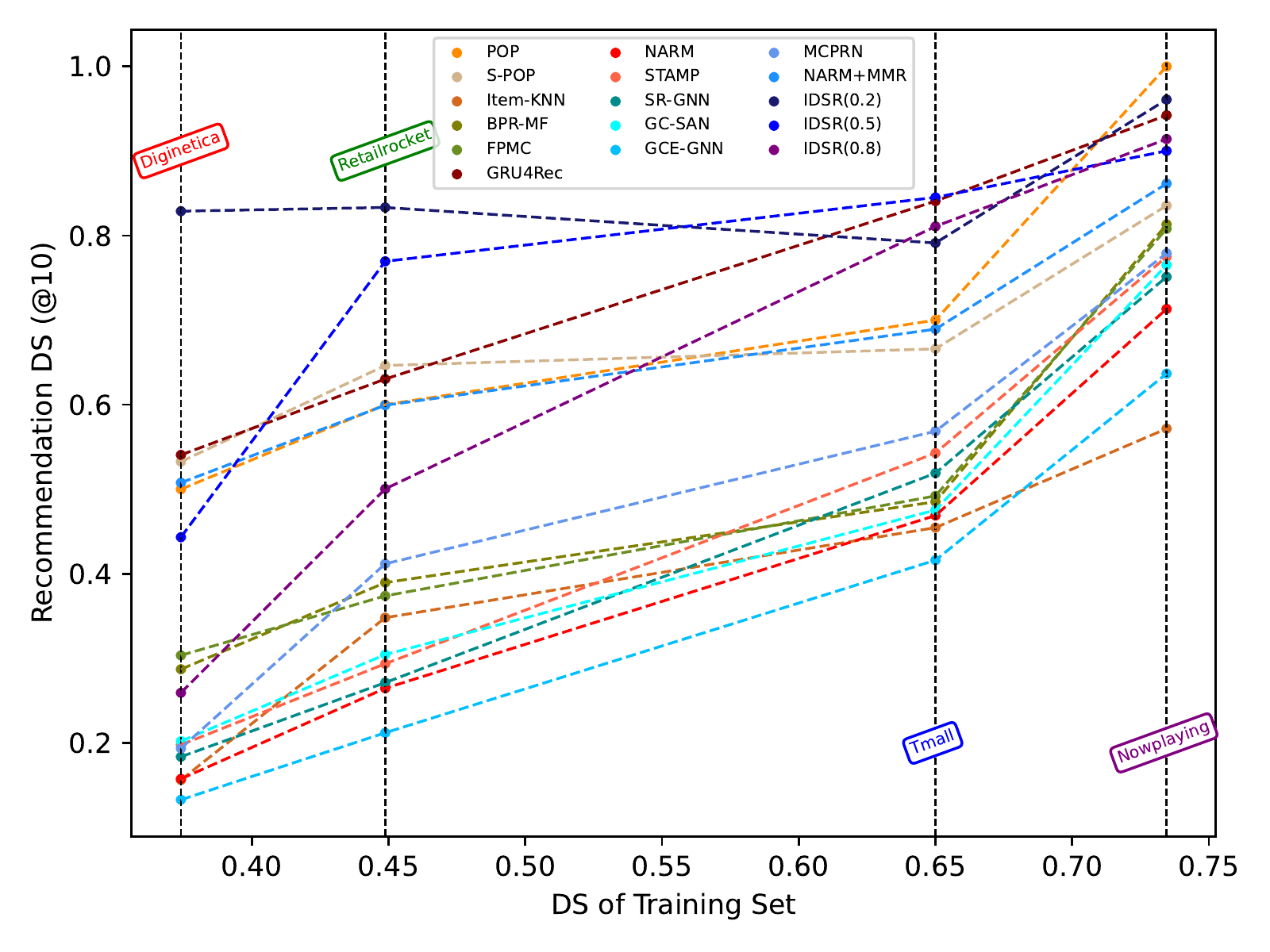}
    \caption{The Influence of Session Diversity of Datasets.}
    \label{fig:dsinfluential}
\end{figure}

\begin{figure}
    \centering
    \includegraphics[scale=0.7]{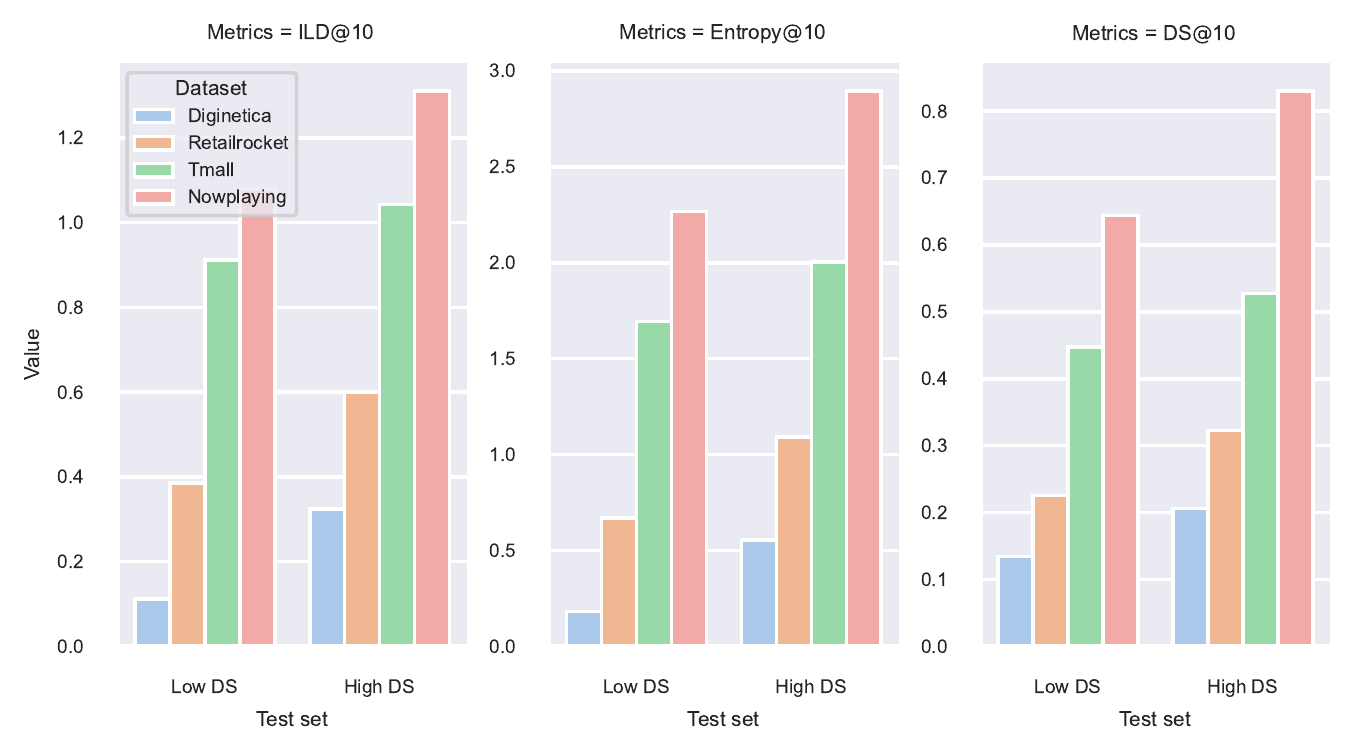}
    \caption{{The Impact of Session Diversity of Test Set on NARM.}}
    \label{fig:dsimpact_test}
\end{figure}

\begin{figure*}
    \centering
     \subfloat[The Influence on ILD]{
         \label{fig:lenild}
         \includegraphics[width=0.8\linewidth]{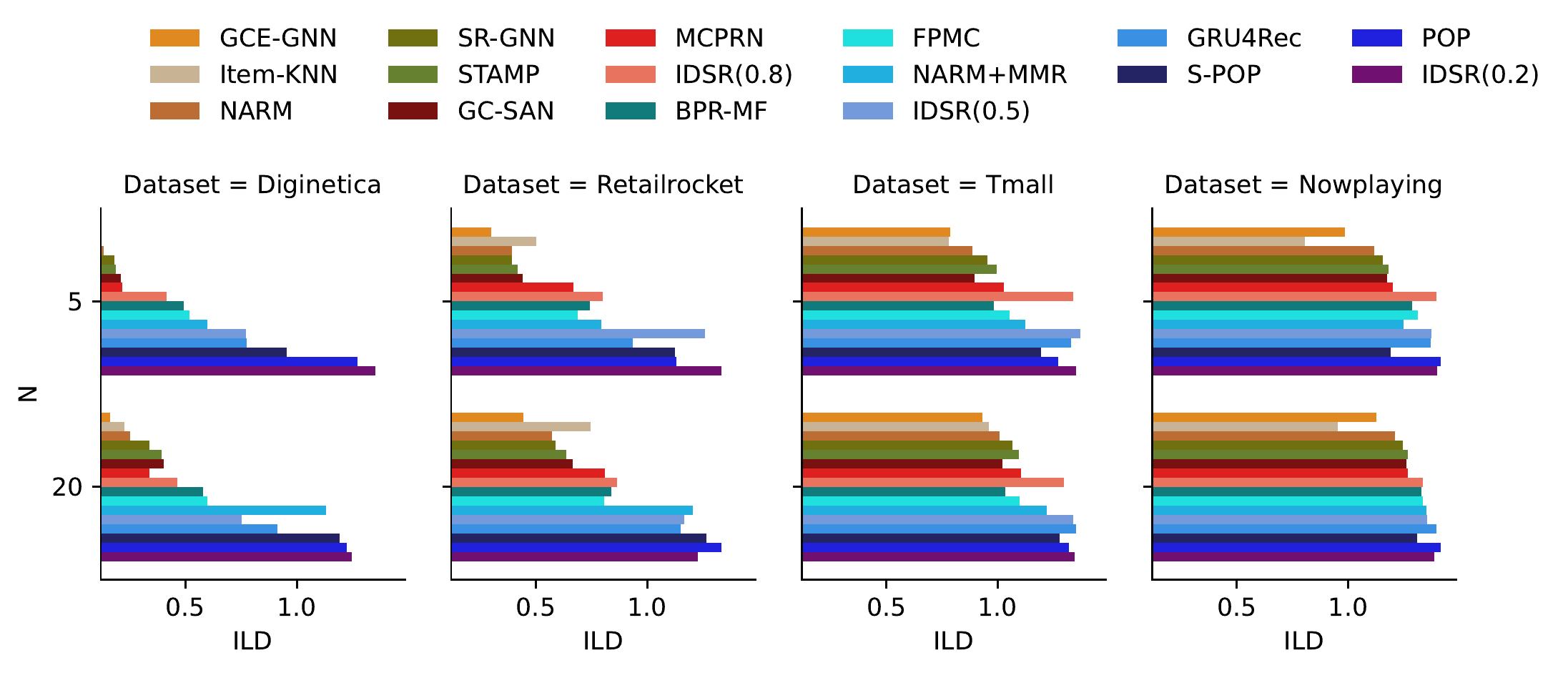}}\\
     \subfloat[The Influence on Entropy]{
         \label{fig:lenentropy}
         \includegraphics[width=0.8\linewidth]{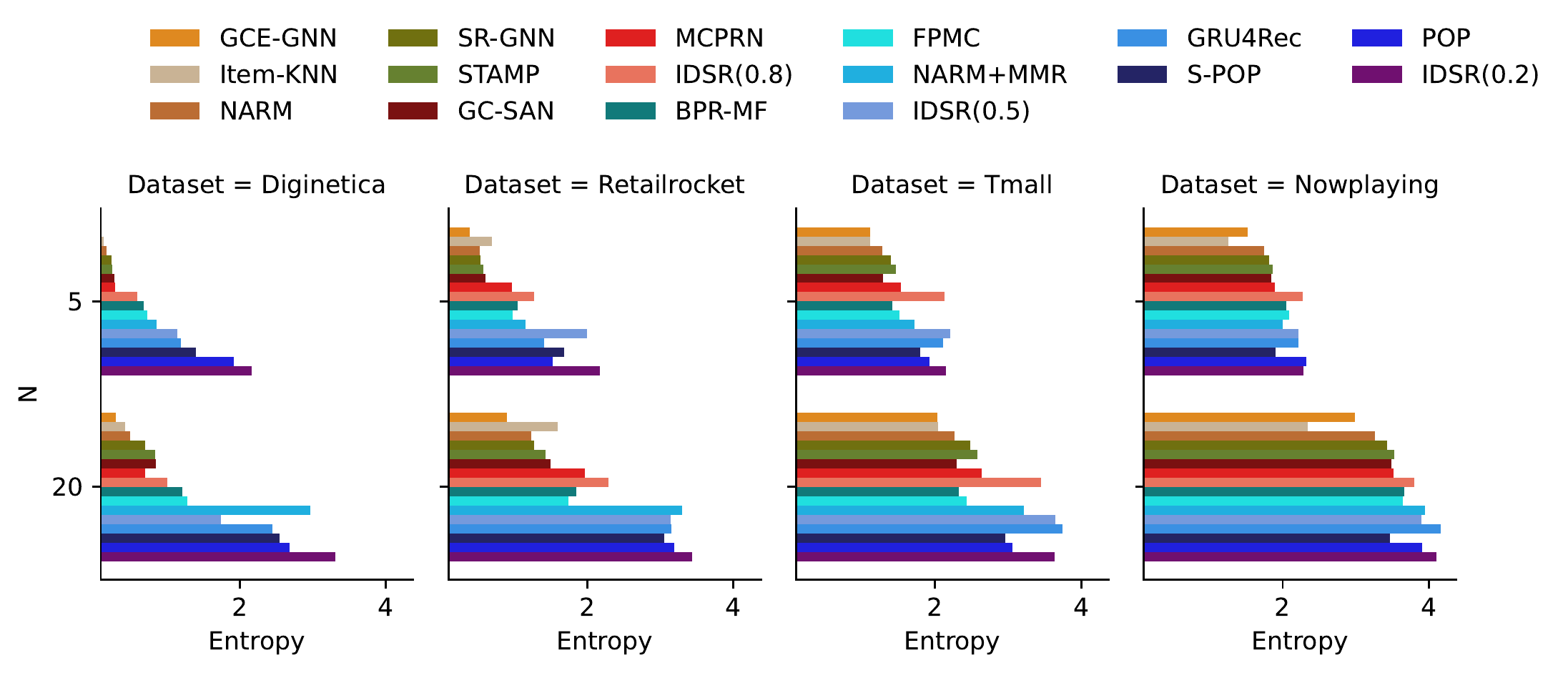}}\\
     \subfloat[The Influence on DS]{
         \label{fig:lends}
         \includegraphics[width=0.8\linewidth]{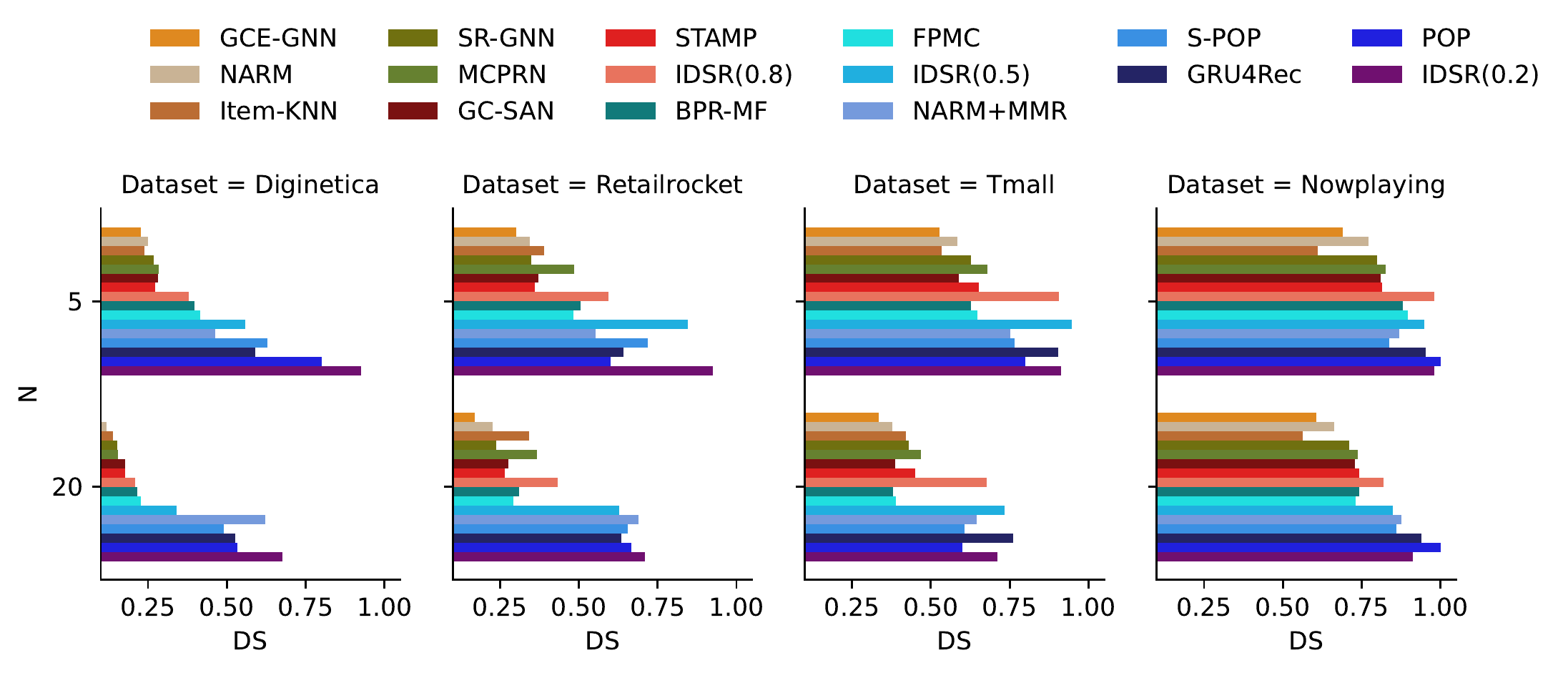}}
    \caption{The Influence of Length of Recommendation Lists.}
    \label{fig:lenbias}
\end{figure*}
\subsection{Influential Factors of Diversity (RQ3)}
Based on the above analysis, we seek to further identify the possible influential factors, besides the complex model designs, that could improve diversity in SBRSs, with the goal of providing guidance towards better diversified SBRSs. In particular, we mainly discuss three factors: granularity of item categorization, session diversity of datasets, and length of recommendation lists. Additionally, we attempt to provide an intuitive idea regarding model designs mainly based on the in-depth analysis regarding learned item embeddings by different types of approaches in Section \ref{sec:in-depth}.

\subsubsection{Granularity of Item Categorization}
Most of the popular diversity metrics (e.g., ILD and Entropy) are calculated via item category information. In this view, towards the same Top-$N$ recommendation list, higher diversity is inclined to be obtained on finer granularity of item categorization, that is, having a larger number of categories and more levels of hierarchy. 

As shown in Table~\ref{tab:datasetStatistics}, the number of categories on Diginetica, Retailrocket, Tmall, and Nowplaying is $995$, $944$, $822$, and $11,558$ respectively. Meanwhile, as can be seen in Tables~\ref{tab:digi}-\ref{tab:nowplaying}, the diversity performance gaps among baselines are decreasing as the number of categories increases across the four datasets (see Figure \ref{fig:dsinfluential}, and its explanation is deferred to Section~\ref{subsubsec:dp}). For example, w.r.t DS@$10$, the variance of all baselines is $0.0360$ on Diginetica, but $0.0133$ on Nowplaying. It implies that the improvement on the performance gap (w.r.t. DS@10) of worse-performing method (e.g., GCE-GNN) over that of better one (e.g., IDSR($\lambda=0.2$)) on Nowplaying ($\lvert 0.6367 - 0.9608 \rvert = 0.3241$) compared to that on Diginetica ($\lvert 0.1328 - 0.8127 \rvert = 0.6799$) is attributed to the finer-grained item category instead of the model design per se.
Therefore, it should be kept alert that performance improvement of methods (measured via category-based diversity metrics) on finer-granularity scenario does not necessarily guarantee a better model, and user-perceived diversity~\cite{allen2008perceived} is recommended to be involved in diversified recommendation studies.

\subsubsection{Session Diversity of Datasets}\label{subsubsec:dp}
We plot the diversity performance (i.e., DS@10) of each baseline on every of the four datasets in Figure~\ref{fig:dsinfluential}, where the x-axis represents DS of the corresponding dataset's training set. The DS of training set on Diginetica, Retailrocket, Tmall, and Nowplaying is $0.3741$, $0.4488$, $0.6500$, and $0.7345$ respectively, and a larger value means that user sessions are more diverse regarding corresponding training set. {Additionally, the DS of test set on above four datasets is $0.3721$, $0.4724$, $0.6278$, and $0.7998$, respectively, which is nearly identical to its DS value on the corresponding training dataset and maintains the same ordering.} It should be noted that, since Diginetica, Retailrocket, and Tmall have similar number of categories, the aforementioned granularity level of item categorization on diversity performance can be conditionally ignored. 
From Figure~\ref{fig:dsinfluential}, we can conclude that the diversity performance is positively correlated with the session diversity of input datasets, which is consistent across baselines. This suggests that a model inclines to generate more diversified recommendation for historically more diversified sessions. The results are quite similar to personalized diversity according to every historical session~\cite{wu2019recent}. {
Furthermore, we also investigate how the DS (diversity score) value of a test set affects final results. To do so, we divide each test set into two categories, Low DS and High DS, using NARM as an example. We evaluate the diversity performance (ILD, entropy, and DS) of the NARM on each group. Our analysis reveals that the diversity of the test samples is positively correlated with recommendation diversity performance, indicating that a more diverse test set results in better performance regarding diversity.}

{
The rationale of the above phenomenon is that the SOTA SBRSs generally determine the relevance among items based on their co-occurrence and sequential relationship. 
Even without item category information exploited in SBRSs, the relevance learned by SBRSs between items belonging to different categories may be considerable if the co-occurrence of these items is significant. Therefore, with higher DS value of the training set, strong relevance between items belonging to different categories could be learned via SBRSs, which ultimately helps produce more diversified recommendation lists. Accordingly, the test samples with higher DS values contain items from more different categories. Based on the relevance among items learned from training set, items that are related to historical items belonging to various categories will be chosen and added into the recommendation list, thus leading to higher diversity performance.
}

\subsubsection{Length of Recommendation Lists}

Figures~\ref{fig:lenild}-\ref{fig:lends} plot the diversity performance of every baseline in terms of ILD, Entropy and DS respectively, regarding varied length of recommendation lists ($N=\{5, 20\}$).
 
First, we can see that from low diversity scenario (Diginetica) to high diversity one (Nowplaying), for every baseline, its diversity performance with regard to each diversity metric increases, further validating the positive correlation between diversity performance and session diversity of datasets (as discussed in Section~\ref{subsubsec:dp}). 
Second, when $N=10\rightarrow 20$, the diversity performance w.r.t. ILD (or Entropy) for every baseline is consistently increasing as depicted in Figure~\ref{fig:lenild}, whereas that regarding DS decreases in Figure~\ref{fig:lends}.
This implies that ILD (or Entropy) is positively associated with the length of the recommendation list $N$ within a model, while that on DS is on the opposite, that is, negatively correlated with $N$. This {can} be caused by, that 
when $N$ increases, more unique categories are likely to occur and thus pair-wise diversity metric (ILD) and category distribution (Entropy) also tend to grow. 
However, DS removes the bias from length of recommendation list by dropping its effect (with appropriate design).
In this case,
DS will decrease if the increasing rate of new categories is lower than that of $N$.

In conclusion, ILD and Entropy suffers from the length bias of recommendation list, whereas DS moderately address this issue. Since in real scenarios, user sessions are mostly of different lengths, diversity metrics like DS are more suitable than those like ILD and Entropy for capturing diversity preference of variable-length sessions. That is to say, we may consider to design diversity-related objective aligned with DS-style metrics.

\subsubsection{A Model Design Guideline for Diversified SBRSs}

It is widely known that diversity is vitally important in traditional recommendation, the same goes for SBRSs. Most existing studies seek for increasingly complex and advanced deep neural structures to improve recommendation accuracy in SBRSs, while keeping a better balance between the two goals remains a challenging problem. It is definitely unconvincing and unacceptable to blindly improve diversity while greatly sacrificing accuracy.
Here we attempt to provide an intuitive idea for tackling this challenge, from the perspective of examining item embeddings distribution grouped by different categories.
As shown in Figure~\ref{fig:umap_digi_20} and discussed in Section~\ref{sec:in-depth}, the representative SBRSs (e.g., NARM and GCE-GNN) with satisfying recommendation accuracy can learn closely connected embeddings of items from the same category. On the other hand, while obtaining better diversity, the distance between embeddings of items from different categories is also reduced by, for example, using RNN-based structure or MLP.
It should be noted that for these representative non-diversified deep methods, they do not explicitly consider the category information.
Therefore, we argue that, for more effective model designs, it is promising to exert appropriate constraints on learned item embeddings, e.g., asking for distinguished (for accuracy) yet diverse (for diversity) item embeddings regarding categories, to obtain better comprehensive performance.

\begin{figure*}[htb]
	\centering
	\begin{tikzpicture}
	\begin{groupplot}[group style={
		group name=myplot,
		group size= 3 by 1,  horizontal sep=1.0cm}, height=4.5cm, width=5cm,
	ylabel style={yshift=-0.65cm},
	legend style = {font=\tiny, column sep=-0.5cm},
	every tick label/.append style={font=\tiny}
	]
    \hspace{-0.2in}
	\nextgroupplot[ybar=0.10,
	bar width=0.4em,
	ylabel={HR@10},
	scaled ticks=false,
	yticklabel style={/pgf/number format/.cd,fixed,precision=3},
 	ymin=0, 
	enlarge x limits=0.25,
	symbolic x coords={Diginetica, Retailrocket, Tmall, Nowplaying},
	ylabel style = {font=\tiny},
	xtick=data,
	]
	\addplot[color=teal, fill=teal, opacity=0.7] coordinates {
		(Diginetica, 0.5018) (Retailrocket, 0.4945) (Tmall, 0.0336) 
            };\label{plots:plot1}
	\addplot[color=pink, fill=pink, opacity=0.7] coordinates {
		(Diginetica, 0.4869) (Retailrocket, 0.5053) (Tmall, 0.0385) 
            };\label{plots:plot2}
	\addplot coordinates {
		(Diginetica, 0.4764) (Retailrocket, 0.5011) (Tmall, 0.0393) 
            };\label{plots:plot3}
  
	\nextgroupplot[ybar=0.1,
	bar width=0.4em,
	ylabel={ILD@10},
	scaled ticks=false,
	yticklabel style={/pgf/number format/.cd,fixed,precision=3},
	ymin=0, 
	enlarge x limits=0.25,
	symbolic x coords={Diginetica, Retailrocket, Tmall, Nowplaying},
	xtick=data,
	ylabel style =  {font=\tiny}]
	\addplot[color=teal, fill=teal, opacity=0.7] coordinates {
		(Diginetica, 0.2704) (Retailrocket, 0.5313) (Tmall, 1.0449) 
            };
	\addplot[color=pink, fill=pink, opacity=0.7] coordinates {
		(Diginetica, 0.1764) (Retailrocket, 0.3852) (Tmall, 0.9484) 
            };
	\addplot coordinates {
		(Diginetica, 0.3326) (Retailrocket, 0.5625) (Tmall, 1.0080) 
            };

  	\nextgroupplot[ybar=0.1,
	bar width=0.4em,
	ylabel={F-score@10},
	scaled ticks=false,
	yticklabel style={/pgf/number format/.cd,fixed,precision=3},
	ymin=0, 
	enlarge x limits=0.25,
	symbolic x coords={Diginetica, Retailrocket, Tmall, Nowplaying},
	xtick=data,
	ylabel style =  {font=\tiny}]
       
	\addplot[color=teal, fill=teal, opacity=0.7] coordinates {
		(Diginetica, 0.1381) (Retailrocket, 0.2530) (Tmall, 0.0292) 
            };
	\addplot[color=pink, fill=pink, opacity=0.7] coordinates {
		(Diginetica, 0.0864) (Retailrocket, 0.1864) (Tmall, 0.0309) 
            };
	\addplot coordinates {
		(Diginetica, 0.1545) (Retailrocket, 0.2745) (Tmall, 0.0334) 
            };

	\end{groupplot}
 
	\path (myplot c1r1.north west|-current bounding box.center)--
	coordinate(legendpos)
	(myplot c3r1.north east|-current bounding box.north);
	\matrix[
	matrix of nodes,
	anchor=south,
	row 1/.style = {nodes={font=\tiny}},
	draw,
	inner sep=0.1em,
	]at([yshift=1.2cm]legendpos)
	{
		\ref{plots:plot1} STAMP&
		\ref{plots:plot2} STAMP+PCP&
		\ref{plots:plot3} STAMP+NCP\\
    };
	\end{tikzpicture}
	\caption{Effects of PCP and NCP on STAMP in terms of HR, ILD, and F-score metrics $N=10$.}\label{fig:cpd_result_10}
\end{figure*}
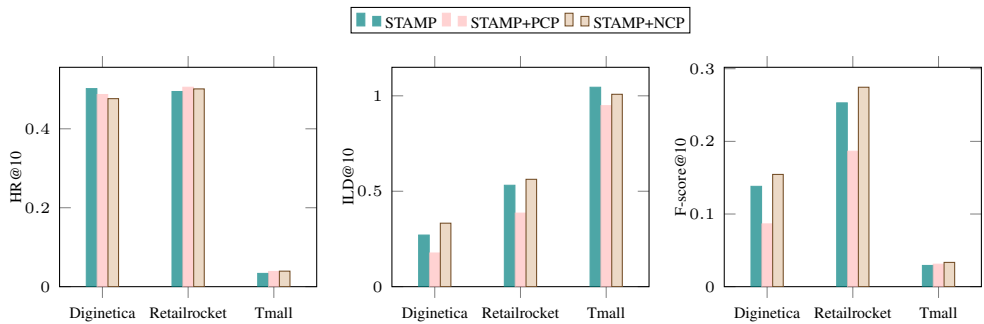

{
Inspired by the aforementioned discovery, we propose a potential solution and conduct a demo experiment to showcase its effectiveness.
For the session-based recommendation, we create a \emph{category prototype} by averaging the learned embeddings of items belonging to the same category. Similarly, we build a \emph{session prototype} by averaging the embeddings of items that occur in the session.
By calculating the Euclidean distance between the session prototype and the corresponding category prototype, we determine the probability of the subsequent category, with a shorter distance resulting in a higher likelihood. 
To create a supervised signal for the next-category target, we develop a next-category cross-entropy loss named $\mathcal{L}_{proto}$. 
We combine this loss function with the next-item prediction loss function $\mathcal{L}_{item}$ from typical SBRSs in two ways:
(1) the overall loss function $\mathcal{L}_{item}+\mathcal{L}_{proto}$ aims to capture more precise preferences with an additional category target; and (2) the overall loss function $\mathcal{L}_{item}-\mathcal{L}_{proto}$ encourages embeddings from different categories to mix and be inseparable. We refer to these two combinations as the Projection Constraint Plugin (abbr. PCP) and the Normalization Constraint Plugin (abbr. NCP), respectively. The PCP plugin projects embeddings from different categories to different subspaces, making them easier to distinguish, while the NCP plugin enforces a normalization constraint that encourages mixing of the embeddings from different categories. To test the effectiveness of our PCP and NCP on e-commerce datasets (i.e., Diginetica, Retailrocket, and Tmall), we use STAMP as the foundation of our SBRS and combine it with these two aforementioned components. The experimental results are displayed in Figure~\ref{fig:cpd_result_10}.}

\begin{figure}[htb]
    \centering
    \includegraphics[scale=0.62]{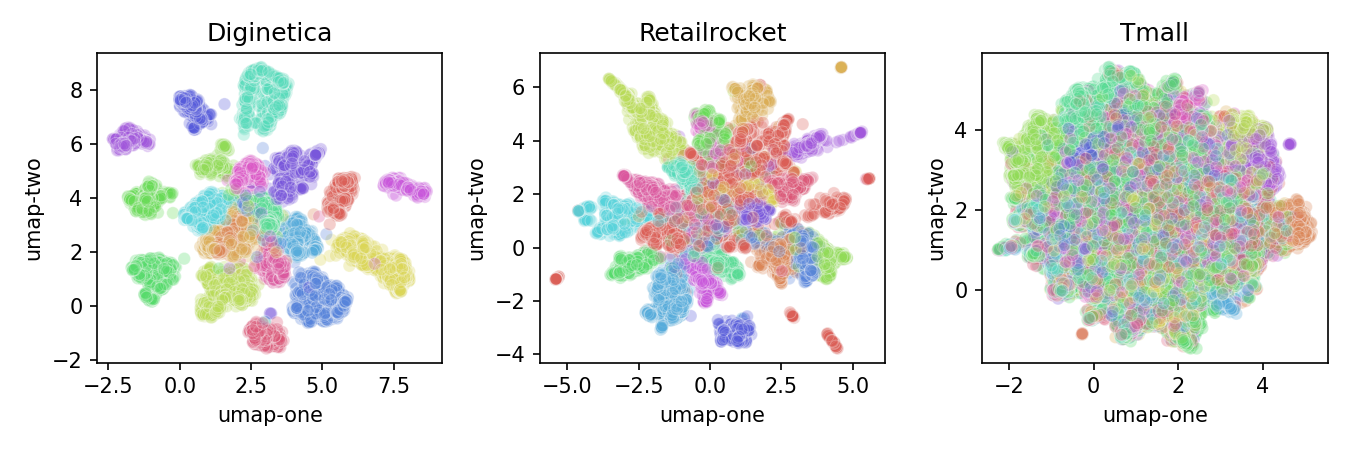}
    \caption{{STAMP's Reduced Dimensional Embeddings of Items from the Most Popular $20$ Categories on E-commerce datasets.}}
    \label{fig:stamp_umap_ecom}
\end{figure}

{
As observed in Figure~\ref{fig:cpd_result_10}, PCP has shown higher accuracy (HR) compared to NCP on Diginetica and Retailrocket. However, in terms of diversity (ILD) and comprehensive performance (F-score), NCP outperforms PCP on all three datasets. These results support the claims made by both NCP and PCP. Additionally, our PCP and NCP exhibit higher accuracy on Retailrocket and Tmall compared to STAMP. In terms of diversity, NCP outperforms STAMP on Diginetica and Retailrocket. Furthermore, NCP outperforms STAMP in comprehensive performance on all three datasets. However, our PCP and NCP fail to outperform STAMP in HR on Diginetica and ILD on Tmall respectively, which is primarily attributed to the dataset's properties. The degree of diversity in the dataset influences the learning of item embeddings. The training set's DS is $0.3741$, $0.4488$, and $0.6500$ for Diginetica, Retailrocket, and Tmall, respectively, where a higher value implies more diversified user sessions. For instance, Figure~\ref{fig:stamp_umap_ecom} depicts the gradual blending of item embeddings (learned by STAMP) of different categories from Diginetica to Tmall. While Diginetica's item embeddings from different categories are separable, Tmall's item embeddings cannot be distinguished. As a result, our PCP does not offer any additional accuracy assistance and may even lead to overfitting on Diginetica, while our NCP does not provide additional diversity support on Tmall.}

\section{Conclusion}
Towards better understanding on diversified recommendation, we have conducted extensive experiments to evaluate {16} state-of-the-art SBRSs with regard to accuracy, diversity and comprehensive performance on four representative datasets.
Our experimental findings show that, for accuracy, deep neural methods perform significantly better than traditional non-neural methods, where GCE-GNN ranks the first place. For diversity, IDSR performs consistently well, proving the effectiveness of its diversity module. 
Meanwhile, non-diversified methods, POP, S-POP and GRU4Rec also gain satisfying performance w.r.t. diversity metrics, implying that non-diversified methods can still maintain a promising diversity performance.
In addition, we provide an in-depth analysis to explore the underlying reasons leading to the varied performance of different deep neural methods using a case study. We find that the representative SBRSs with encouraging recommendation accuracy can obtain closely connected embeddings of items from the same category, while models with more diverse item embeddings can obtain better diversity.
Our empirical analysis also unveil that the relationship between accuracy and diversity
is quite complex and mixed. Besides the ``trade-off'' relationship,
they {can} be positively correlated with each other, that is, having a same-trend (win-win or lose-lose) relationship, which does exist across different methods and datasets.
We have also identified three possible influential factors, besides the complex model design, that can be capable of improving diversity in SBRSs: granularity of item categorization, session diversity of datasets, and length of recommendation lists. We further offer an intuitive idea for better model-designs based on the relationships of item embeddings of different categories.
{Furthermore, in order to aid understanding of the intuitive guideline, we strive to offer a practical solution and carry out a demonstration experiment to illustrate its efficacy.}
For future study, we plan to design advanced diversified methods for session-based recommendation according to our findings.

\newpage
\appendix
\section{Additional Results for Section~\ref{sec:in-depth}}

\begin{figure}[htbp]
\centering
    \includegraphics[scale=0.41]{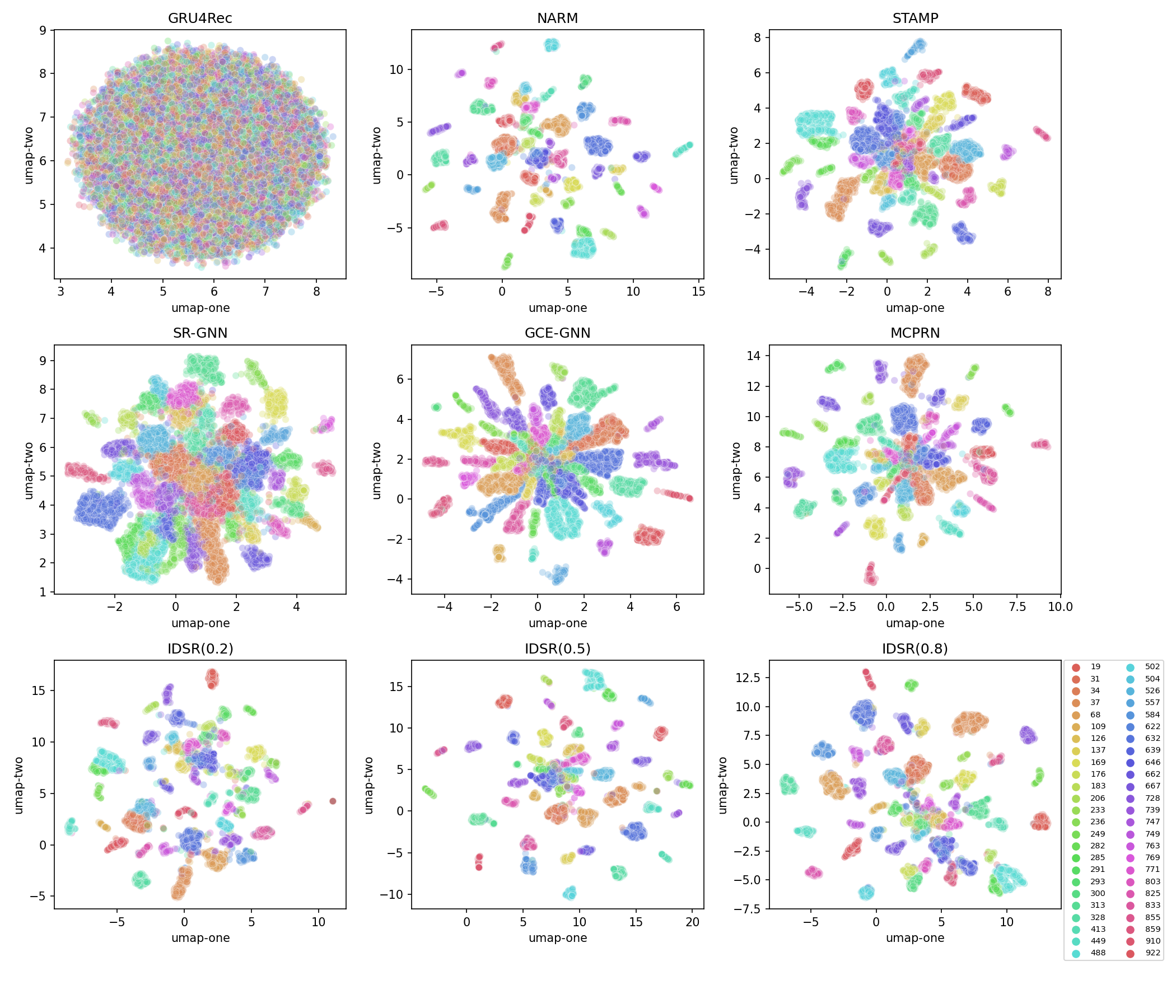}
    \caption{{Reduced Dimensional Embeddings (Using UMAP) Of Items From the Most Popular $50$ Categories.}}
    \label{fig:umap_digi_50}
\end{figure}

\begin{figure}[htbp]
\centering
    \includegraphics[scale=0.35]{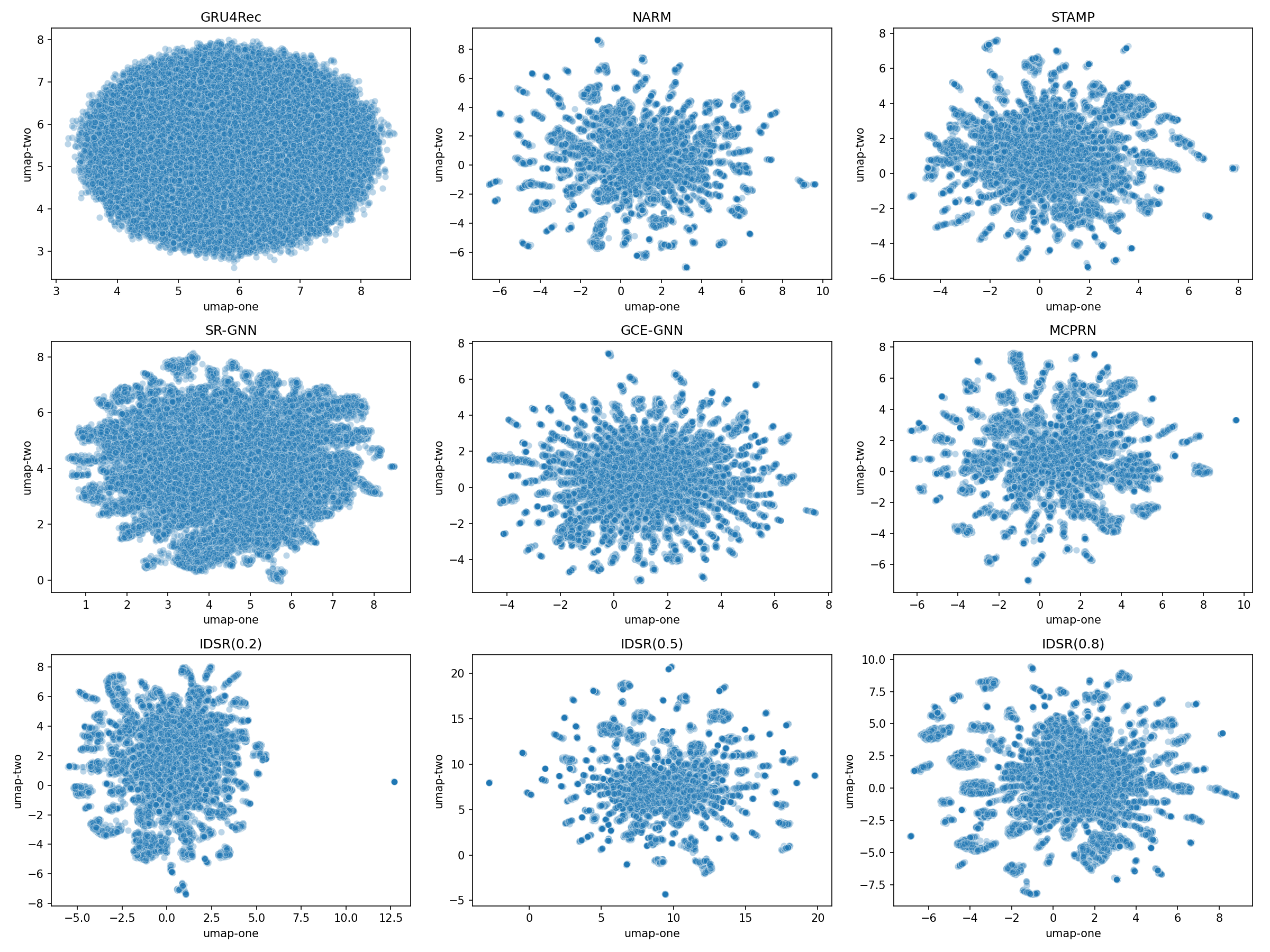}
    \caption{{Reduced Dimensional Embeddings (Using UMAP) Of All Items.}}
    \label{fig:umap_digi_total}
\end{figure}

\section{Additional Results for Section~\ref{sec:acc-div-rel}}
\begin{figure}[htbp]
\centering
    \includegraphics[scale=0.295]{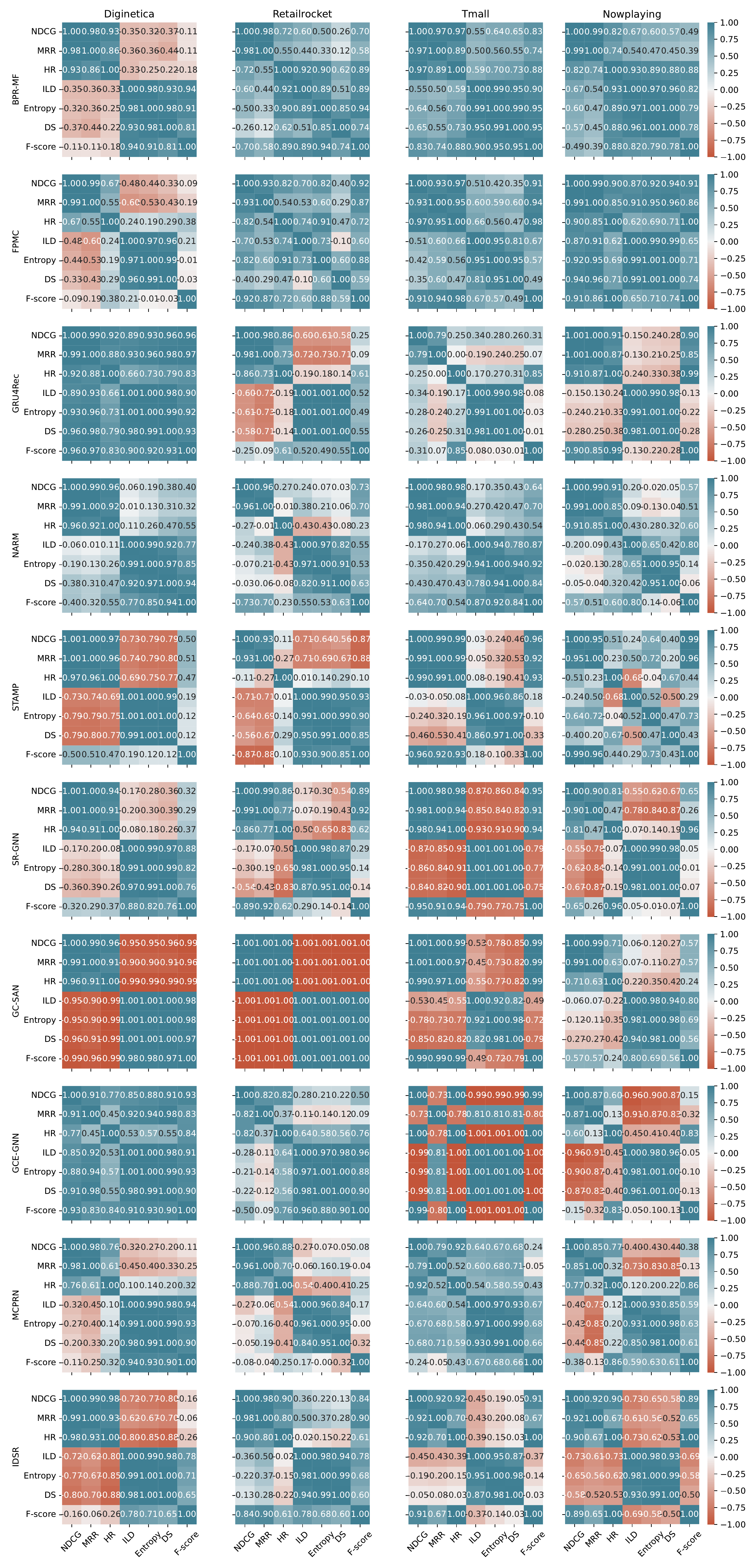}
    \vspace{-4mm}
    \caption{{Pearson Correlation Coefficient of Metrics for Different Baselines on Every Dataset. Each value is calculated given two arrays by concatenating Top-10 performance (running 5 times) of each baseline on each dataset.}}
    \label{fig:corre_bydataandmodel_total}
    \vspace{-0.1in}
\end{figure}

\bibliographystyle{unsrt}  
\bibliography{references}

\end{document}